\documentclass[a4paper,11pt]{article}

\usepackage{amsmath}
\usepackage{amsthm}
\usepackage{physics}
\usepackage{amsfonts}
\usepackage{graphicx}
\usepackage{amssymb}
\usepackage{caption}
\usepackage{multicol}
\usepackage{multirow}
\usepackage{booktabs}
\usepackage[normalem]{ulem}
\usepackage[margin=2.2cm]{geometry}
\usepackage[final]{hyperref} 
\usepackage{cite}
\def\linkcolor{cyan!70!black}
\hypersetup{
	colorlinks=true,
	linkcolor=\linkcolor,
	citecolor=\linkcolor,
	filecolor=\linkcolor,
	urlcolor=\linkcolor         
}
\usepackage[capitalize]{cleveref}
\usepackage[dvipsnames]{xcolor}
\begin{document}

\begin{titlepage}
\vspace*{-1cm}
\phantom{hep-ph/***}
\flushright 
CPPC-2025-04

\vskip 1.5cm
\begin{center}
\mathversion{bold}
{\LARGE\bf
{Selection rules for charged lepton flavour violating\\[0.1cm] 
processes from residual flavour groups
\mathversion{normal}
}}
\vskip .3cm
\end{center}
\vskip 0.5  cm
\begin{center}
{\large Lorenzo~Calibbi}$\,^{a}$, 
{\large Claudia~Hagedorn}$\,^{b}$,
{\large Michael~A.~Schmidt}$\,^{c}$, and
{\large James~Vandeleur}$\,^{c}$
\\
\vskip .7cm
{\footnotesize
$^{a}$ School of Physics, Nankai University, Tianjin 300071, China\\[1mm]
$^{b}$ Instituto de F\'isica Corpuscular, Universitat de Val\`encia -- CSIC,\\
Parc Cient\'ific, Catedr\'atico Jos\'e Beltr\'an 2, 46980 Paterna, Spain\\[1mm]
$^{c}$ Sydney Consortium for Particle Physics and Cosmology, School of Physics,\\ The University of New South Wales, Sydney, New South Wales 2052, Australia
\vskip .5cm
\begin{minipage}[l]{.9\textwidth}
\begin{center}
\textit{E-mail:}
\href{mailto:calibbi@nankai.edu.cn}{\tt{calibbi@nankai.edu.cn}},
\href{mailto:claudia.hagedorn@ific.uv.es}{\tt{claudia.hagedorn@ific.uv.es}},\\
\href{mailto:m.schmidt@unsw.edu.au}{\tt{m.schmidt@unsw.edu.au}},
\href{mailto:james.vandeleur@unsw.edu.au}{\tt{james.vandeleur@unsw.edu.au}}
\end{center}
\end{minipage}
}
\end{center}
\vskip 1cm

\begin{abstract}
We systematically investigate the possible phenomenological impact of residual flavour groups in the charged lepton sector. We consider all possible flavour charge assignments for
abelian residual symmetries up to $\mathbb{Z}_8$.
The allowed flavour structures of operators in Standard Model Effective Field Theory (up to dimension six) lead to distinctive and observable patterns of charged lepton flavour violating processes. We illustrate the relevance of such selection rules displaying the current bounds on and the future sensitivities to the new physics scale. These results demonstrate, in particular, the importance and discriminating power of searches for lepton flavour violating $\tau$ lepton decays and muonium to antimuonium conversion. 
\end{abstract}

\end{titlepage}
\setcounter{footnote}{0}

\section{Introduction}
\label{sec:intro}

The Standard Model (SM) has been very successful in describing and precisely accounting for a wide range of phenomena in particle physics and cosmology. Nevertheless, it has several short-comings that call for its extension: fermion masses and mixing can only be adjusted, but not explained, a mechanism
for generating neutrino masses is missing, the baryon asymmetry of the Universe cannot be correctly accommodated, and a viable Dark Matter candidate is absent.

Discrete flavour symmetries (also in combination with a CP symmetry) have been successfully applied to the lepton sector in particular in order to explain the observed fermion mixing, see e.g.~\cite{King:2015aea,Feruglio:2019ybq,Xing:2020ijf,Kobayashi:2022moq,Grimus:2011fk} for reviews. A particularly predictive 
approach assumes that in the charged and neutral lepton sector different residual symmetries, $G_\ell$ and $G_\nu$, are preserved, whose mismatch corresponds to lepton mixing. Such residual symmetries are abelian in general, 
since non-abelian groups would impose too strong constraints (for example, two exactly degenerate fermion masses). 
Well-known examples are the flavour symmetries $A_4$ and $S_4$ that can predict tri-bimaximal lepton mixing, see e.g.~\cite{Altarelli:2005yx,He:2006dk,Lam:2008rs}, if the residual symmetries are $G_\ell=\mathbb{Z}_3$ and $G_\nu=\mathbb{Z}_2 \times \mathbb{Z}_2$.\footnote{In the case of $A_4$ the second $\mathbb{Z}_2$ symmetry is accidental and not contained in the original flavour group.} The observation of the reactor mixing angle being of the size of the Cabibbo angle, 
$\sin^2 \theta_{13}\approx 0.022$~\cite{Esteban:2024eli}, requires modifications to this result and numerous dedicated studies can be found in the literature, see for instance~\cite{Lin:2008aj,Shimizu:2011xg,King:2011ab,deMedeirosVarzielas:2012cet}. 

Assuming the existence of a residual symmetry $G_\ell$ among charged leptons, it is interesting to consider its further phenomenological consequences. As has already been pointed out~\cite{Feruglio:2008ht,Csaki:2008qq,Ma:2010gs,Cao:2011df,Hagedorn:2011un,Holthausen:2012wz,Pascoli:2016wlt,Muramatsu:2016bda,Bigaran:2022giz,Lichtenstein:2023iut,Lichtenstein:2023vza}, the residual $\mathbb{Z}_3$ symmetry, usually called lepton triality, constrains possible signals of charged lepton flavour violation (cLFV). In $A_4$ and $S_4$ models, the charged lepton flavours, $e$, $\mu$ and $\tau$, typically 
transform differently under $G_\ell=\mathbb{Z}_3$, e.g.~$e \sim 1$, $\mu \sim \omega$ and $\tau \sim \omega^2$ with $\omega$ being the third root of unity which corresponds to $e$ having flavour charge $0$, $\mu$ charge $1$, and $\tau$ charge $2$ modulo $3$. Then, radiative cLFV processes such as $\mu\to e \gamma$ are forbidden in the limit of an exact residual symmetry, while the two tri-lepton decays $\tau\to ee\bar{\mu}$ and $\tau\to\mu\mu\bar{e}$ are compatible with $G_\ell=\mathbb{Z}_3$. 

This example shows how residual symmetries can lead to the existence of certain selection rules that leave an observable imprint on the pattern of cLFV processes. This observation is of particular interest given that cLFV is a general prediction of extensions of the SM that address neutrino masses and other unexplained phenomena, while at the same time it is subject to stringent experimental constraints, see e.g.~\cite{Calibbi:2017uvl,Ardu:2022sbt,Davidson:2022jai}. The existence of selection rules could entail that the discovery of cLFV signals does not occur through the processes most searched for. On the other hand, this highlights the power of cLFV searches to discriminate among different residual flavour symmetries and different new physics models more generally. There is the exciting possibility that present and future experiments are capable of unveiling the underlying organising principle in the charged lepton sector. For earlier discussions along these lines\,---\,not necessarily related to flavour symmetries\,---\,see e.g.~\cite{Blankenburg:2012nx,Jones-Perez:2013uma,Heeck:2016xwg,Heeck:2024uiz,Greljo:2022cah,Palavric:2024gvu,Heeck:2025jfs}. 

In this work, we analyse cLFV processes compatible with different residual symmetries $G_\ell$ among the charged leptons and different possible flavour charge assignments of the flavours $e$, $\mu$ and $\tau$. 
Concretely, we study all cyclic groups $\mathbb{Z}_N$ with $N=2,\dots, 8$ and all possible flavour charge combinations, including those for which two of the lepton flavours transform in the same way. First, we systematically construct and classify all possible flavour structures (with up to 16 charged lepton fields) that are compatible with the imposed residual symmetry. Then, we apply these flavour structures to SM Effective Field Theory (SMEFT) operators~\cite{Buchmuller:1985jz,Grzadkowski:2010es,Henning:2014wua,Brivio:2017vri,Isidori:2023pyp} and derive phenomenological implications and constraints. In a first step, we work out which cLFV processes,
considering up to dimension-six SMEFT operators, are allowed and compare the results among different residual symmetries $G_\ell=\mathbb{Z}_N$, highlighting which patterns of observables could provide evidence for a certain $G_\ell$ and, potentially, also flavour charge assignment. We only include operators that conserve (total) lepton number, $\Delta L=0$, since, in general, both the mechanism and the scale of lepton number violation can differ from those of (charged) lepton flavour violation.
In a second step, we set the coefficients of all permitted operators
to values motivated by different types of ultraviolet (UV) theories, derive lower limits on the new physics scale $\Lambda$ using existing experimental bounds, and show the future reach in $\Lambda$ based on prospective sensitivities. This enables us to  assess the role of different cLFV observables in testing the discussed selection rules within the context of the present and future experimental landscape.

The remainder of this paper is organised as follows. In Section~\ref{sec:motivation} we give details about the motivation of this study. Section~\ref{sec:theory} is dedicated to the systematic listing of the flavour structures allowed for the different groups $G_\ell$ and the different flavour charge assignments. 
Section~\ref{sec:classifyops} comprises the results for allowed and forbidden cLFV processes for the different residual symmetries $G_\ell=\mathbb{Z}_N$.
In Section~\ref{sec:boundsLambda} we use these results and make various assumptions regarding the size of the coefficients of the allowed SMEFT operators, in order to estimate present and future limits on the new physics scale $\Lambda$. 
We summarise and give a brief outlook in Section~\ref{sec:summ}. The appendices contain technical details: a demonstration of the completeness of the study (Appendix~\ref{app:flavProof}), the list of the flavour structures for $G_\ell$ larger than $\mathbb{Z}_5$ (Appendix~\ref{app:allowedOps}), all possible flavour charge assignments for $G_\ell=\mathbb{Z}_3$ (Appendix~\ref{app:expandedNotation}), as well as the present and future experimental limits employed in the analysis and numerical bounds on $\Lambda$ 
(Appendices~\ref{app:scale} and~\ref{app:scale2}).

\section{Motivation}
\label{sec:motivation}

In this section, we motivate different choices of the residual symmetry $G_\ell$ and comment on possible flavour charge assignments. 

Typically, the discrete groups employed as flavour symmetries (and thus also their subgroups) are subgroups of $SU(3)$, see~\cite{Grimus:2011fk}. This constrains the possible combinations of flavour charges, as their sum must equal zero 
modulo $N$ for $G_\ell=\mathbb{Z}_N$. In this study, we do not impose this constraint but do highlight the instances in which the flavour charge assignment can correspond to the action of $\mathbb{Z}_N$ being represented by a special unitary matrix, see Section~\ref{sec:chargeNotation}.

When explaining fermion mixing as the mismatch of two distinct residual symmetries of a (discrete) flavour group, 
 the predictive power is maximised if the three fermion generations are distinguished by their charge under $G_\ell=\mathbb{Z}_N$, since otherwise at least one free parameter enters the mixing matrix. Indeed, the choice $G_\ell=\mathbb{Z}_3$
 is the minimal one that can ensure the absence of such a free parameter. We do not impose this as a constraint in the following, and consequently, we include $G_\ell=\mathbb{Z}_2$ in the list of residual symmetries and also permit combinations of flavour charges for which two lepton flavours have the same charge.
 
The group $G_\ell=\mathbb{Z}_3$ is not only the minimal possible choice for maximising the predictive power regarding fermion mixing, it is also frequently the residual symmetry in the charged lepton sector when considering different
flavour symmetries and deriving lepton mixing. Indeed, many viable lepton mixing patterns can be obtained from members of the series $\Delta (3 \, n^2)$~\cite{Luhn:2007uq} and $\Delta (6 \, n^2)$~\cite{Escobar:2008vc}, potentially combined with CP, where 
$G_\ell=\mathbb{Z}_3$, while the residual symmetry $G_\nu$ is a Klein group (or the direct product of $\mathbb{Z}_2$ and CP); for studies on lepton mixing see e.g.~\cite{King:2013vna,Ding:2015rwa,Hagedorn:2014wha,Ding:2014ora}.
 
An example with $G_\ell=\mathbb{Z}_4$ can be obtained from the flavour group $S_4$, where, together with $G_\nu=\mathbb{Z}_2 \times \mathbb{Z}_2$, it gives rise to bimaxmial  mixing~\cite{deAdelhartToorop:2011re}.
 Similarly, for the flavour symmetry $A_5$ (and CP) the residual symmetries $G_\ell=\mathbb{Z}_5$ and $G_\nu=\mathbb{Z}_2 \times \mathbb{Z}_2$ (or $G_\nu=\mathbb{Z}_2 \times CP$) 
can lead to the so-called golden ratio mixing~\cite{Feruglio:2011qq} (or golden ratio-type, if a CP symmetry is also involved~\cite{DiIura:2015kfa,Ballett:2015wia,Li:2015jxa}). The residual symmetry $G_\ell=\mathbb{Z}_7$ instead can play a role
in scenarios with the flavour group $PSL (2,7)$ (also called $\Sigma (168)$), see e.g.~\cite{deAdelhartToorop:2011re}. 
Further examples of residual symmetries $G_\ell$ are encountered e.g.~in the analysis of the flavour groups $\Sigma (n \, \varphi)$~\cite{Hagedorn:2013nra}.
Hence, it is well-motivated to study the phenomenology of residual symmetries in the charged lepton sector other than lepton triality. Still, residual symmetries $G_\ell=\mathbb{Z}_N$ with $N$ small usually occur and it is, thus, reasonable 
to only consider $N \leq 8$. 

One may also employ direct products of cyclic symmetries as $G_\ell$. However, they do not lead to new results as the impact of $G_\ell=\mathbb{Z}_2 \times \mathbb{Z}_2$ can be deduced from the combination
of the results for $G_\ell=\mathbb{Z}_2$, and similarly for $G_\ell=\mathbb{Z}_2 \times \mathbb{Z}_2 \times \mathbb{Z}_2$ as well as $G_\ell=\mathbb{Z}_2 \times \mathbb{Z}_4$; see comments in Section~\ref{sec:results}. The group $\mathbb{Z}_2 \times \mathbb{Z}_3$ is isomorphic to $\mathbb{Z}_6$ and therefore the results corresponding to this choice are also covered in the present study.

As happens in models implementing the idea that the mismatch of the residual symmetries $G_\ell$ and $G_\nu$ encodes lepton mixing, none of the residual symmetries are exact. In general, $G_\ell$ is broken, e.g.~by (small) shifts in the alignment of the vacuum of the flavour symmetry breaking fields as well as by contributions of higher-dimensional operators to the charged lepton sector that involve fields that break the original flavour symmetry to $G_\nu$, 
see e.g.~\cite{Altarelli:2005yx}. Consequently, we expect that a process forbidden by $G_\ell$ is not strictly absent in a concrete model, but rather has a signal strength that is (highly) suppressed compared to those of processes 
that are compatible with $G_\ell$. In the case of lepton triality, for example, a signal of $\tau\to \mu\mu\bar{\mu}$ should be suppressed compared to $\tau\to ee\bar{\mu}$\,---\,see e.g.~\cite{Pascoli:2016wlt}. 

Eventually, we point out that in concrete realisations the branching ratios of two processes compatible with $G_\ell$ may have different sizes, e.g.~the branching ratio of $\tau\to \mu\mu\bar{e}$ is much larger than that of $\tau\to ee\bar{\mu}$ in the supersymmetric version of the well-known $A_4$ model~\cite{Muramatsu:2016bda}. Nevertheless, this conclusion depends on whether the flavour symmetry breaking fields mix or not,
as shown in~\cite{Pascoli:2016wlt}. Such a mixing can be achieved via the inclusion of certain soft supersymmetry breaking terms.

In the current study, we do not consider explicit models and thus we do not take into account corrections to the breaking of the residual symmetry $G_\ell$ in the charged lepton sector, nor effects that can induce a hierarchy among the signal strengths of different processes allowed by $G_\ell$. Rather, we focus on the possible constraints on cLFV processes arising from $G_\ell$ different from $\mathbb{Z}_3$ and for all possible flavour charge assignments.

\section{Flavour charge assignments and flavour structures}
\label{sec:theory}

We begin by systematically listing the flavour structures of operators with charged lepton fields which are compatible with the (non-trivial) flavour charge assignments for the different choices of $G_\ell$.\footnote{We assume that only charged leptons and no other SM particles are charged under $G_\ell$.} 
With this at hand, we construct in Section~\ref{sec:classifyops} all allowed SMEFT operators up to dimension six and categorise the cLFV processes that are permitted by a given $G_\ell$ and flavour charge assignment. Thus, in this section, the Lorentz and other structures of the operators are suppressed so that only their lepton flavour structure is specified. For example, we have the following correspondences
\begin{align}
        (\bar{\ell}_{\mu}\gamma_{\nu}\ell_{e})(\bar{\ell}_{\mu}\gamma^{\nu}\ell_{e}) \to ee\mu^{\dag}\mu^{\dag} \ \ \mbox{and} \ \ (\varphi^{\dag}\varphi)(\bar{\ell}_{\tau}e_{e}\varphi) \to e\tau^{\dag} \, ,
\end{align}
where $\ell_i$ is a left-handed lepton doublet of flavour $i$, $e_j$ a right-handed charged lepton of flavour $j$, $\varphi$ denotes the Higgs field, and $e$, $\mu$ and $\tau$ are the three lepton flavours.

\subsection{Constraints on number of charged leptons and flavour charges}
\label{sec:LNconstraints}

Consider an operator that conserves the number of charged leptons and can contain any number of the three flavours of the charged lepton fields. The {\it individual charged lepton numbers} must obey the condition
\begin{align}
\label{eq:echargeconstraint}
        (n_e^- - n_e^+) + (n_\mu^- - n_\mu^+) + (n_\tau^- - n_\tau^+) &\equiv \Delta n_e + \Delta n_\mu + \Delta n_\tau = 0 \ , 
\end{align}
where $n^-_{e,\mu,\tau}$, $n_{e,\mu,\tau}^+$ are respectively the number of particle and antiparticle fields of charged lepton flavour $e$, $\mu$ and $\tau$ and $\Delta n_{e,\mu,\tau}$ is the change in the corresponding individual charged lepton number.

Similarly, for $e$, $\mu$ and $\tau$ carrying flavour charge $\alpha$, $\beta$ and $\gamma$ under $G_\ell=\mathbb{Z}_N$, the conservation of {\em lepton flavour charge} imposes 
the condition
\begin{align} 
\label{eq:fchargeconstraint}
        \alpha\, \Delta n_e + \beta \,\Delta n_\mu + \gamma \, \Delta n_\tau &= 0 \mod N  \ .
\end{align}
These flavour charges are defined modulo $N$.
Any operator with individual charged lepton numbers satisfying~\cref{eq:echargeconstraint,eq:fchargeconstraint} is permitted by the residual symmetry $G_\ell$ and the flavour charge assignment.

\subsection{Notation of flavour charge assignments} 
\label{sec:chargeNotation}

We represent an assignment of lepton flavour charges under $G_\ell=\mathbb{Z}_N$ as 
\begin{equation}
\label{eq:ZNabc}
\mathbb{Z}_N(\alpha,\beta,\gamma) \, ,
\end{equation}
with $\alpha$, $\beta$ and $\gamma$ being the flavour charges for $e$, $\mu$ and $\tau$, respectively. Using the two constraints in~\cref{eq:echargeconstraint,eq:fchargeconstraint}, we can reduce the labeling of the flavour charge assignment from three to two independent parameters.
An allowed flavour structure must satisfy~\cref{eq:fchargeconstraint}
\begin{align}
        0 \mod N &= \alpha \, \Delta n_{e} + \beta \, \Delta n_{\mu} + \gamma \, \Delta n_{\tau} = \alpha \, (\Delta n_{e} + \Delta n_{\mu} + \Delta n_{\tau}) + \delta_{1} \Delta n_{\mu} + (\delta_{1} + \delta_{2}) \Delta n_{\tau} \ , \label{eq:fChargeReduction}
\end{align}
with $\delta_{1} \equiv \beta-\alpha$ and $\delta_{2} \equiv \gamma - \beta$ defined as the differences between the assigned flavour charges.
Taking into account the conservation of the number of charged leptons, see~\cref{eq:echargeconstraint}, we arrive at
\begin{align}
\label{eq:Cfconstraint}
        0 \mod N &= \delta_{1} \Delta n_{\mu} + (\delta_{1} + \delta_{2}) \Delta n_{\tau}\ .
\end{align}
This leaves only a constraint on the parameters $\delta_{1}$ and $\delta_{2}$, while also being independent of $\Delta n_{e}$. Note that the particular permutation of the charged lepton flavours is arbitrary. However, one of the differences of individual charged lepton numbers is determined in terms of the other two by means of conservation of charged lepton number, see Eq.~(\ref{eq:echargeconstraint}). Therefore, we introduce the following reduced notation 
\begin{equation}
\label{eq:reducednotation}
N (\delta_1, \delta_2) \, .
\end{equation}
It can be derived from the notation used in Eq.~(\ref{eq:ZNabc}), e.g.~for $\mathbb{Z}_3(0,1,2)$ we have 
\begin{align}
\delta_{1} = \beta - \alpha = 1-0 = 1 \ , \ \
        \delta_{2} = \gamma - \beta = 2-1 = 1 \ ,
\end{align}
leading to 
\begin{align}
    \mathbb{Z}_3(0,1,2) \to 3(1, 1) \ .
\end{align}
Importantly, the reduced notation, in general, does not uniquely specify a flavour charge assignment. For example, several of the latter correspond to $3(0,1)$, i.e.
\begin{align}
\left.
        \begin{aligned}
            \mathbb{Z}_3(0,0,1)&\\
            \mathbb{Z}_3(1,1,2)&\\
            \mathbb{Z}_3(2,2,0)&
        \end{aligned}\right\} \to 3(0,1) \ .
\end{align}
Although the reduced notation in this example represents three different flavour charge assignments, they all permit the same flavour structures of operators.

If the three flavour charges sum to zero modulo $N$, the action of the $\mathbb{Z}_N$ symmetry on the charged leptons can be represented by a special unitary matrix, see~\Cref{sec:motivation}. From this point onward, we highlight these instances by displaying the flavour charge assignment in boldface. For example, $\mathbf{4(0,1)}$ corresponds to $\mathbf{\mathbb{Z}_4(1,1,2)}$. Note, however, that $\mathbf{4(0,1)}$ also comprises the flavour charge assignment $\mathbb{Z}_4(0,0,1)$, whose flavour charges do not sum to zero modulo four. The reduced notation cannot separate these cases. Indeed, the constraint on the flavour charges for summing to zero is
\begin{align}
\label{eq:suconstraint}
        3 \, \alpha + 2 \, \delta_{1} + \delta_{2} = 0 \mod N \ , 
\end{align}
which not only depends on the parameters $\delta_1$ and $\delta_2$, but also explicitly on the flavour charge of the electron, $\alpha$.
Unless three divides $N$, $3 \, \alpha$ can take any value modulo $N$ and thus there is always an assignment of the three lepton flavour charges that sum to zero modulo $N$.
 
\subsection{Equivalences and redundancies in notation}
\label{sec:notationTransformations}

In the case two or more flavour charge assignments permit the same flavour structures, we consider these assignments equivalent.
We have already shown that the reduced notation $N(\delta_1, \delta_2)$ represents a set of equivalent flavour  charge assignments. However, there are further possibilities for such an equivalence. In the following, 
our choice of conventions is emphasised in \emph{italics}.

\paragraph{Common factors.} Inspecting the constraint in~\cref{eq:Cfconstraint},
 it is clear that applying a common scaling factor to $\delta_{1}$, $\delta_{2}$ and $N$ does not change the constraint. Consequently, two flavour charge assignments allow the same flavour structures, if they are related by
        \begin{align}
           N'(\delta_{1}',\delta_{2}') \leftrightarrow A \, N(A \, \delta_{1},A \, \delta_{2})    
        \end{align}
with $A$ being a rational number. 

The constraint in~\cref{eq:Cfconstraint} is also invariant under the scaling of just $\delta_{1}$ and $\delta_{2}$ by a non-zero integer $B$. 
 The important difference between these two scalings is that $A$ is always invertible and thus there is a (two-way) equivalence between the flavour charge assignments. 
 On the contrary, $B$ can only be inverted in particular cases, meaning the implication only applies in one direction in general. For example, 
   \begin{align}                
   \mathbf{4(0,2)} = \mathbf{4(}2\times \mathbf{0,}\, 2\mathbf{\times 1)}
\end{align}
with $B=2$ implies that flavour structures allowed by the flavour charge assignment $\mathbf{4(0,1)}$ are also allowed by $\mathbf{4(0,2)}$. However, there is no requirement for all flavour structures permitted by $\mathbf{4(0,2)}$ to be allowed by $\mathbf{4(0,1)}$. In particular, the flavour structure
        \begin{align}
                ee\tau^{\dagger}\tau^{\dagger} \text{ is allowed by } & \mathbb{Z}_4(0,0,2) \to \mathbf{4(0,2)} \ ,\nonumber\\
                \text{but not by } &\mathbb{Z}_4(0,0,1) \to \mathbf{4(0,1)} \, .
        \end{align}
Two flavour charge assignments related by a scaling of $\delta_{1}$ and $\delta_{2}$ are only equivalent in cases where this scaling can be inverted, while taking into account the definition of the flavour charges modulo $N$. Indeed, this inversion is always possible for $N$ prime. For $N$ not prime, the scaling can only be inverted, if $B$ and $N$ are coprime.
For example, the flavour charge assignments $\mathbf{5(1,1)}$ and $\mathbf{5(2,2)}$ allow the same set of flavour structures (with $B=2$ and the inverse scaling factor $B^\prime=3$), as do $\mathbf{8(0,1)}$ and $\mathbf{8(0,3)}$ (with $B=3$ and $B^\prime=3$); however, as argued, $\mathbf{4(0,1)}$ and $\mathbf{4(0,2)}$ do not.
    \emph{We always use the smallest values of $N$, $\delta_1$ and $\delta_2$ that represent a set of equivalent flavour charge assignments.}
        
\paragraph{Permutations.} The choice of the labels $e$, $\mu$ and $\tau$ is somewhat arbitrary. Although, for example, $\mathbf{\mathbb{Z}_4(0,1,3)}$ $\to \mathbf{4(1,2)}$ and  $\mathbf{\mathbb{Z}_4(1,3,0)}$ $\to \mathbf{4(2,1)}$ lead to different sets of allowed flavour structures, they only differ by a permutation of the charged lepton flavours, i.e.~$\{e,\mu,\tau\}$ should be traded for $\{\tau,e,\mu\}$. Where they are not important, we suppress permutations of the flavour charges.  
\emph{By convention, we denote the flavour charge assignments with lowest lexicographic ordering, i.e.~with $\delta_{1} \leq \delta_{2}$.}

\paragraph{Hermitian and complex conjugation.}        If an operator with a given flavour structure is allowed, its Hermitian conjugate must be as well, implying a sign change of the changes in individual charged lepton numbers $\Delta n_{e,\mu,\tau}$.
This amounts to an overall sign change in~\cref{eq:echargeconstraint,eq:fchargeconstraint}, but does not alter the equalities. So, whenever we list the allowed flavour structures for a certain flavour charge assignment, it should be understood that {\em operators arising from Hermitian conjugation are implicitly included}.

At the same time, it is clear that using the complex conjugate of the flavour charges $\alpha$, $\beta$ and $\gamma$, i.e.~$N-\alpha$, $N-\beta$, $N-\gamma$, leads to the same allowed flavour structures.

\subsection{Systematic study of flavour structures}
\label{sec:construct}

In the following, we systematically study all flavour structures that are allowed for the different flavour charge assignments and for different $G_\ell=\mathbb{Z}_N$. 

For each $G_\ell$, all possible flavour charge assignments are given by $\mathbb{Z}_N(\alpha,\beta,\gamma)$, where $0 \leq \alpha,\beta,\gamma \leq N-1$. However, it suffices to only search for the allowed flavour structures of a single permutation of the flavour charges $\alpha$, $\beta$, $\gamma$ and then 
permute the labels of the charged leptons in order to obtain the results for the remaining permutations. Thus, we can restrict the search to\footnote{Using the reduced notation, see Eq.~(\ref{eq:reducednotation}), it is sufficient to take into account flavour charge assignments with $0 \leq \delta_1\leq \delta_2$ and $0 \leq \delta_2 \leq N-1$. For $G_\ell=\mathbb{Z}_N$ with $N \leq 8$, there is little difference in computation time to scan over either set of assignments.}
\begin{equation}
\label{eq:rangesabc}
0 \leq \alpha \leq \beta \ , \ \ 
0 \leq \beta \leq \gamma \ , \ \
0 \leq \gamma \leq N-1 \; .
\end{equation}
In the next step, we consider the flavour structures. Each flavour structure corresponds to a list of the changes in individual charged lepton numbers, $\lbrace\Delta n_{e}, \Delta n_{\mu}, \Delta n_{\tau}\rbrace$. Ignoring trivial flavour structures, such as $e e^{\dagger}$, since these do not have an effect on the change in individual lepton number, a (negative) positive value of $\Delta n_{e,\mu,\tau}$ corresponds to a net number of \mbox{(anti)particles}. For example, $\lbrace\Delta n_{e}, \Delta n_{\mu}, \Delta n_{\tau}\rbrace= \lbrace2,-1,-1\rbrace$ encodes the flavour structure $e e\mu^{\dagger}\tau^{\dagger}$. We check for each flavour charge assignment the parameter space of flavour structures that is spanned by
\begin{equation}
\label{eq:rangesDeltana}
     0 \leq   \Delta n_{e} \leq N \ , \ \ -N \leq \Delta n_{\mu} \leq N \ , \ \ \Delta n_{\tau} = -\Delta n_{e} - \Delta n_{\mu} 
\end{equation}
and filter those flavour structures that satisfy the conservation of flavour charge, see~\cref{eq:fchargeconstraint}. The conservation of charged lepton number is ensured by the choice of $\Delta n_\tau$. 
 A few comments are in order: despite the flavour charge being defined modulo $N$, we include both $0$ and $N$ in the ranges in~\cref{eq:rangesDeltana}, as the individual charged lepton numbers are  counted without modulo $N$. For example, with $ \mathbb{Z}_2(0,0,1)$, the flavour structure $ee\tau^\dag\tau^\dag$, corresponding to $\Delta n_e = 2$, $\Delta n_\mu = 0$, and $\Delta n_\tau = -2$, is allowed and cannot be 
 reduced to/deduced from $e\tau^\dag$, since the latter is not permitted. 
 Allowed flavour structures with values of $\Delta n_{e,\mu,\tau}$ being outside of the ranges in~\cref{eq:rangesDeltana} can be constructed via Hermitian conjugation (for negative $\Delta n_{e}$) or arise from a combination of flavour structures with smaller values of the changes in individual charged lepton numbers (for $\abs{\Delta n_{e,\mu,\tau}} > N$). The latter is shown in~\Cref{app:flavProof}.

\begin{table}[tb!]
        \renewcommand\arraystretch{1}
\begin{tabular}{lcl}
        \toprule
        Flavour &  & Flavour \\
        charges & $d_\ell$ &  structures\\\midrule
\multirow[t]{4}{8em}{$\mathbf{2(0,1)}$}
 & $3$ &$e\mu^\dagger $\\
 & $6$ &$\mu\mu\tau^\dagger \tau^\dagger $\\
 &  &$e\mu\tau^\dagger \tau^\dagger $\\
 &  &$ee\tau^\dagger \tau^\dagger $\\
\midrule
\multirow[t]{5}{8em}{$3(0,1)$}
 & $3$ &$e\mu^\dagger $\\
 & $9$ &$\mu\mu\mu\tau^\dagger \tau^\dagger \tau^\dagger $\\
 &  &$e\mu\mu\tau^\dagger \tau^\dagger \tau^\dagger $\\
 &  &$ee\mu\tau^\dagger \tau^\dagger \tau^\dagger $\\
 &  &$eee\tau^\dagger \tau^\dagger \tau^\dagger $\\
\midrule
 \multirow[t]{6}{8em}{$\mathbf{3(1,1)}$}
 & $6$ &$e\mu\tau^\dagger \tau^\dagger $\\
 &  &$e\mu^\dagger \mu^\dagger \tau$\\
 &  &$ee\mu^\dagger \tau^\dagger $\\
 & $9$ &$\mu\mu\mu\tau^\dagger \tau^\dagger \tau^\dagger $\\
 &  &$eee\tau^\dagger \tau^\dagger \tau^\dagger $\\
 &  &$eee\mu^\dagger \mu^\dagger \mu^\dagger $\\
 \midrule
 \multirow[t]{6}{8em}{$\mathbf{4(0,1)}$}
 & $3$ &$e\mu^\dagger $\\
 & $12$ &$\mu\mu\mu\mu\tau^\dagger \tau^\dagger \tau^\dagger \tau^\dagger $\\
 &  &$e\mu\mu\mu\tau^\dagger \tau^\dagger \tau^\dagger \tau^\dagger $\\
 &  &$ee\mu\mu\tau^\dagger \tau^\dagger \tau^\dagger \tau^\dagger $\\
 &  &$eee\mu\tau^\dagger \tau^\dagger \tau^\dagger \tau^\dagger $\\
 &  &$eeee\tau^\dagger \tau^\dagger \tau^\dagger \tau^\dagger $\\
\bottomrule
\end{tabular}
\hspace{0.2in}
\begin{tabular}{lcl}
        \toprule
        Flavour &  & Flavour \\
        charges & $d_\ell$ &  structures\\\midrule
\multirow[t]{6}{8em}{$\mathbf{4(1,1)}$}
 & $6$ &$e\mu^\dagger \mu^\dagger \tau$\\
 &  &$ee\tau^\dagger \tau^\dagger $\\
 & $9$ &$e\mu\mu\tau^\dagger \tau^\dagger \tau^\dagger $\\
 &  &$eee\mu^\dagger \mu^\dagger \tau^\dagger $\\
 & $12$ &$\mu\mu\mu\mu\tau^\dagger \tau^\dagger \tau^\dagger \tau^\dagger $\\
 &  &$eeee\mu^\dagger \mu^\dagger \mu^\dagger \mu^\dagger $\\
\midrule
\multirow[t]{7}{8em}{$\mathbf{5(0,1)}$}
 & $3$ &$e\mu^\dagger $\\
 & $15$ &$\mu\mu\mu\mu\mu\tau^\dagger \tau^\dagger \tau^\dagger \tau^\dagger \tau^\dagger $\\
 &  &$e\mu\mu\mu\mu\tau^\dagger \tau^\dagger \tau^\dagger \tau^\dagger \tau^\dagger $\\
 &  &$ee\mu\mu\mu\tau^\dagger \tau^\dagger \tau^\dagger \tau^\dagger \tau^\dagger $\\
 &  &$eee\mu\mu\tau^\dagger \tau^\dagger \tau^\dagger \tau^\dagger \tau^\dagger $\\
 &  &$eeee\mu\tau^\dagger \tau^\dagger \tau^\dagger \tau^\dagger \tau^\dagger $\\
 &  &$eeeee\tau^\dagger \tau^\dagger \tau^\dagger \tau^\dagger \tau^\dagger $\\
\midrule
\multirow[t]{8}{8em}{$\mathbf{5(1,1)}$}
 & $6$ &$e\mu^\dagger \mu^\dagger \tau$\\
 & $9$ &$ee\mu\tau^\dagger \tau^\dagger \tau^\dagger $\\
 &  &$eee\mu^\dagger \tau^\dagger \tau^\dagger $\\
 & $12$ &$e\mu\mu\mu\tau^\dagger \tau^\dagger \tau^\dagger \tau^\dagger $\\
 &  &$eeee\mu^\dagger \mu^\dagger \mu^\dagger \tau^\dagger $\\
 & $15$ &$\mu\mu\mu\mu\mu\tau^\dagger \tau^\dagger \tau^\dagger \tau^\dagger \tau^\dagger $\\
 &  &$eeeee\tau^\dagger \tau^\dagger \tau^\dagger \tau^\dagger \tau^\dagger $\\
 &  &$eeeee\mu^\dagger \mu^\dagger \mu^\dagger \mu^\dagger \mu^\dagger $\\
\bottomrule
\end{tabular}
        \caption{Allowed flavour structures for each inequivalent flavour charge assignment for $G_\ell=\mathbb{Z}_N$ with $N\leq 5$. For further description and details, see text.}
        \label{tbl:allowedOps}
        \centering
\end{table}
\subsection{Results}
\label{sec:results}

We summarise the results of the flavour structures allowed by the flavour charge assignments for $G_\ell=\mathbb{Z}_N$ with $N \leq 8$. Taking into account the equivalences and redundancies in notation outlined in~\Cref{sec:notationTransformations} greatly compacts the list of flavour charge assignments to only the inequivalent ones. For $N \leq 5$, these can be found in~\Cref{tbl:allowedOps} together with their allowed flavour structures. The results for larger $N$ are given in~\Cref{app:allowedOps}. Note that for
 $N\geq5$ the different flavour charge assignments allow for at most a single flavour structure
 with no more than four charged leptons. Given that these correspond to dimension-six SMEFT operators, the study of residual symmetries $G_\ell=\mathbb{Z}_N$ with larger $N$ is phenomenologically less interesting. For this reason, we focus on flavour charge assignments that can be obtained for $N\leq 4$ in Sections~\ref{sec:classifyops} and~\ref{sec:boundsLambda}.   
 
For the sake of brevity, in~\Cref{tbl:allowedOps} we do not include Hermitian conjugates of the flavour structures, trivial flavour combinations (e.g.~$ee^{\dagger}$) nor flavour structures containing such combinations nor flavour structures that arise from combining flavour structures already shown in this table for a given flavour charge assignment. Flavour charge assignments and their corresponding flavour structures that result from a permutation of the labels of the charged leptons of a flavour charge assignment shown in~\Cref{tbl:allowedOps} are also excluded. As mentioned 
in~\Cref{sec:chargeNotation}, flavour charge assignments that comprise combinations of flavour charges which sum up to zero modulo $N$ are indicated in boldface.
It is interesting to note that all but one of the flavour charge assignments shown in~\Cref{tbl:allowedOps}, namely $3(0,1)$, have this property. In~\Cref{app:expandedNotation} we provide the list of all possible flavour charge assignments for $G_\ell=\mathbb{Z}_3$ as an explicit example of the notation.

As indicated in Section~\ref{sec:motivation}, results for residual symmetries $G_\ell$ that are direct products of cyclic groups are not explicitly discussed. Nevertheless, they can be deduced from the information found in~\Cref{tbl:allowedOps}. In general, for a flavour charge assignment of such a direct product, the set of allowed flavour structures can be constructed
  as the intersection of the sets of allowed flavour structures for the flavour charge assignments of each group in the product; potentially, taking into account appropriate permutations of the labels of the charged leptons. Furthermore, one has to include larger flavour structures allowed by the flavour charge assignments of each group. For instance, reading from~\Cref{tbl:allowedOps}, a flavour charge assignment $\mathbf{2(0,1)} \times 3(0,1)$ for $G_\ell=\mathbb{Z}_2 \times \mathbb{Z}_3$ (corresponding to $\mathbb{Z}_6$) allows the flavour structure $e\mu^\dag$ (and those that can be constructed following the above prescription). Additionally, it permits flavour structures with twelve charged leptons made from combinations of the allowed flavour structures with six and four fields, respectively, for each group in the product. Such flavour structures are, indeed, found in~\cref{app:allowedOps} for the flavour charge assignment $6(0,1)$ for $G_\ell=\mathbb{Z}_6$.

In~\Cref{tbl:allowedOps}, we have introduced the quantity $d_\ell$ as 
lepton mass dimension  which is defined as
$d_\ell=(\text{no. charged lepton fields})\times 3/2$. 
In this way, the results for the flavour structures can be more easily applied to SMEFT operators. One can clearly see that the largest value of $d_\ell$ for which a given flavour charge assignment results in non-decomposable flavour structures is $3 \, N$. Moreover, such flavour structures always exist. They contain $N$ copies of flavour charged leptons that therefore trivially conserve flavour charge. For flavour charge assignments with two equal flavour charges, such as $\mathbf{2(0,1)}$ or $\mathbf{5(0,1)}$, the flavour structures with the largest value of $d_\ell$ are the only ones that differentiate among them.

\section{Classification of allowed cLFV processes}
\label{sec:classifyops}

We are now in the position to determine which combination of cLFV processes one can expect from a certain residual symmetry $G_\ell$ and flavour charge assignment of the charged leptons. To do so, we translate the flavour structures studied in~\Cref{sec:theory} into SMEFT operators~\cite{Buchmuller:1985jz,Grzadkowski:2010es,Henning:2014wua,Brivio:2017vri,Isidori:2023pyp}. We choose SMEFT, since it is the most general framework for studying beyond SM effects, mediated by new physics heavier than the electroweak scale. As commented in Section~\ref{sec:motivation}, the group $G_\ell$ in general does not correspond to an exact residual symmetry of the theory. As such, references to the `observation'/`non-observation' of a process in the following discussion should be understood as amplification/suppression of this process in comparison to others. Analogously, the `implication'/`exclusion' of a certain residual symmetry $G_\ell$ and flavour charge assignment has to be interpreted as these being favoured/disfavoured by experimental results.

\subsection{Flavour structures of dimension-six SMEFT operators}

For any realistic observation of cLFV, we are restricted to SMEFT operators at dimension six or lower. The dimension-five Weinberg operator violates lepton number~\cite{Weinberg:1979sa} and thus it is beyond the scope of the present study. Moreover, it does not contribute to cLFV processes other than through the modification of the lepton mixing matrix where it suffers from suppression by the smallness of neutrino masses. Consequently, we work within a theory defined by the effective Lagrangian 
\begin{align}
    \mathcal{L}_\textsc{smeft} = \mathcal{L}_\textsc{sm} + \frac{1}{\Lambda^{2}}\sum_{x} \,C_x\mathcal{O}^{d=6}_{x} \,,
    \label{eq:smeft}
\end{align}
where $\Lambda$ is the new physics scale and $C_x$ denotes the Wilson coefficient (WC) of a given dimension-six operator $\mathcal{O}^{d=6}_{x}$. For simplicity, we assume a common new physics scale $\Lambda$ for all dimension-six operators.

Dimension-six SMEFT operators comprising leptons either contain two or four of them~\cite{Crivellin:2013hpa}.
Typical SMEFT operators with two charged leptons are the dipole operators $(\bar{\ell}_i\,\sigma^{\mu\nu} e_j)\, \varphi \, B_{\mu\nu}$ and $(\bar{\ell}_i\,\sigma^{\mu\nu} e_j)\, \varphi \, W_{\mu\nu}$, with the $U(1)_Y$ and $SU(2)_L$ field strength tensors $B_{\mu\nu}$ and $W_{\mu\nu}$, respectively. Upon electroweak symmetry breaking via the vacuum expectation value, $\langle \varphi \rangle = v/\sqrt2$ and $v \approx 246 \, \mbox{GeV}$, these operators combine into the photon dipole operators $\frac{v}{\sqrt2}(\bar{e}_{Li}\,\sigma^{\mu\nu} e_j) \,F_{\mu\nu}$ with $e_{Li}$ being a left-handed charged lepton of flavour $i$ and $F_{\mu\nu}$ the electromagnetic field strength tensor. At low energy, these operators induce radiative decays of heavier charged leptons into lighter ones, i.e.~the processes $\mu\to e\gamma$, $\tau \to \mu \gamma$ or $\tau \to e\gamma$. 

The operators containing four leptons, such as $(\bar{\ell}_i\gamma_\nu\ell_j)(\bar{\ell}_k\gamma^\nu\ell_l)$, can be divided into three categories depending on their flavour structure.
First, we have non-trivial four-lepton operators. These correspond to the flavour structures labelled with lepton mass dimension $d_\ell=6$ in~\Cref{tbl:allowedOps} and include operators that lead to processes such as $\tau \to ee\bar \mu$, muonium to antimuonium conversion $\rm M\to \rm \overline M$ (that is, a bound state $\rm M = \mu^+e^-$ converting into $\rm \overline M = \mu^- e^+$), and scattering processes such as $ee \to \tau\tau$ (standing for $e^\pm e^\pm \to \tau^\pm\tau^\pm$). Second, there are semi-trivial four-lepton operators with one pair of charged leptons forming a trivial flavour combination (e.g.~$ee^\dagger$). Although they can lead to more complicated processes such as $\tau\to \mu e \bar e$, these operators are equivalent to two-lepton operators in terms of the information they provide about the residual symmetry $G_\ell$ and the flavour charge assignment.
Note that operators with a flavour structure such as $ee\mu^{\dag}\mu^{\dag}$ can be constructed as the duplicate of a two-lepton flavour structure, namely $e\mu^\dagger$, however, they do not necessarily admit the same set of $G_\ell$ and flavour charge assignments, compare Table~\ref{tbl:allowedOps}. Such operators are still considered non-trivial four-lepton operators and studied independently from their two-lepton counterparts.
Finally, there exist trivial four-lepton operators that are constructed from two trivial two-lepton flavour structures such as $e e^\dagger\mu\mu^\dagger$. They are of little interest to the analysis at hand as they carry no information about the residual symmetry $G_\ell$ and the flavour charge assignment of the charged leptons.

\subsection{Implications for \mathversion{bold}\texorpdfstring{$G_\ell$}{Gl}\mathversion{normal} and flavour charge assignments}

\begin{figure}[t!]
    \centering
    \includegraphics[width=\textwidth]{"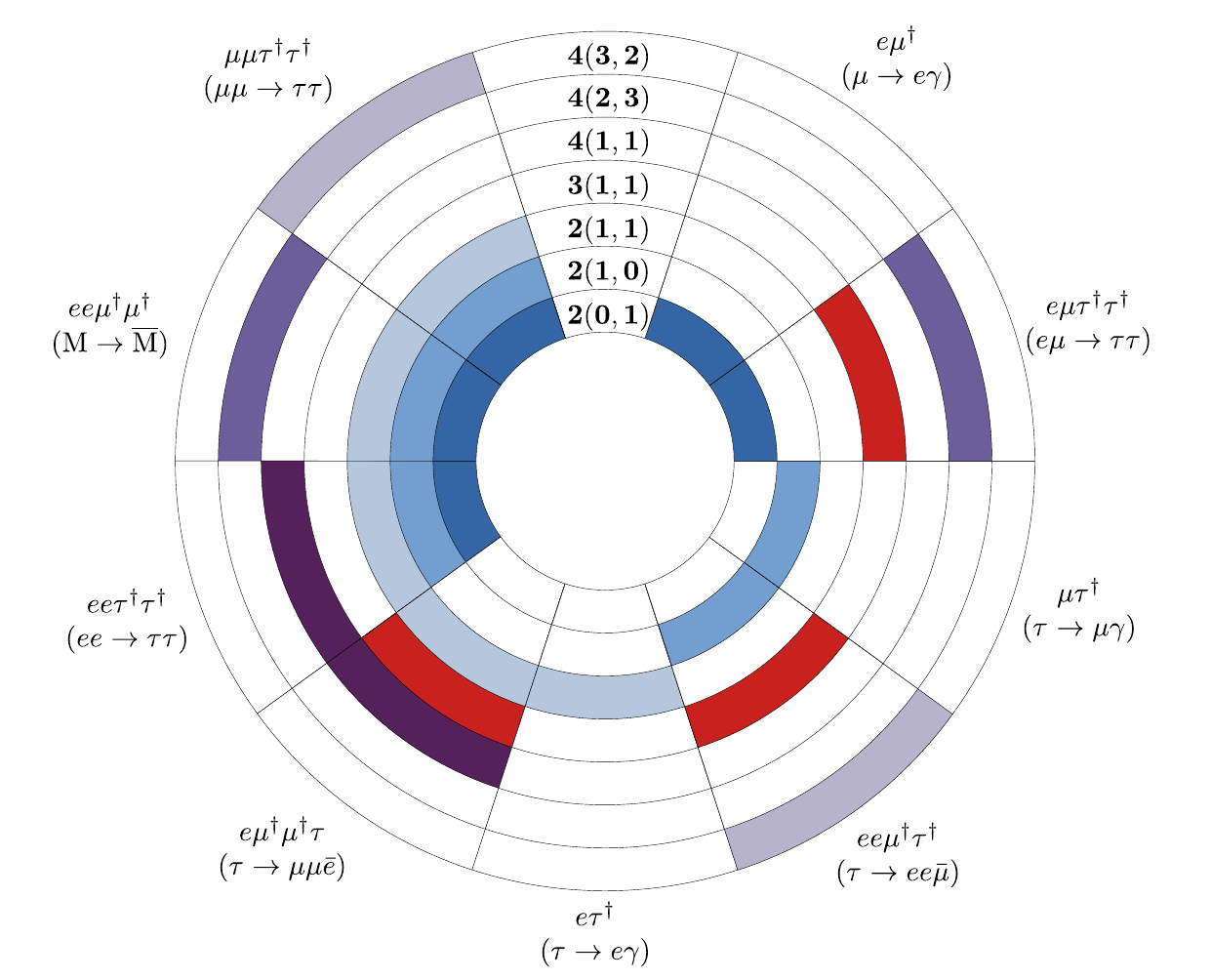"}
    \vspace{0.3cm}\\ 
    {\large Restrictions from single flavour structure}\vspace{0.2cm}\\ 
\begin{tabular}{ccc}
            \hline\vspace{-0.2cm}\\
            $e\mu^\dag$ - $N(0,a)$ & $ee\mu^{\dag}\mu^{\dag}$ - $2N(N,a)$ & $ee\mu^\dag\tau^\dag$ - $N(N-a,2a)$ \\
            $\mu\tau^{\dag}$ - $N(a,0)$ & $\mu\mu\tau^{\dag}\tau^{\dag}$ - $2N(a,N)$ & $e\mu^{\dag}\mu^{\dag}\tau$ - $N(a,a)$\\
            $e\tau^{\dag}$ - $N(a,N-a)$ & $ee\tau^{\dag}\tau^{\dag}$ - $2N(a,N-a)$ & $e\mu\tau^{\dag}\tau^{\dag}$ - $N(2a,N-a)$
    \end{tabular}
    \caption{Illustration of compatibility between flavour charge assignments and flavour structures with $d_\ell \leq 6$, taking into account permutations of the charged lepton flavours. Coloured segments indicate compatibility. 
    Only flavour charge assignments that allow for at least two (independent and non-trivial) flavour structures with up to four leptons, $d_\ell \leq 6$, see~\Cref{tbl:allowedOps}, are shown.
    For each flavour structure an example of a cLFV process is given in parentheses. 
    The table lists restrictions on the flavour charge assignments from observing (at least) one cLFV process corresponding to a certain flavour structure, again accounting for permutations among the charged lepton flavours. The parameter $a$ is an integer.
  }
    \label{fig:rotary}
\end{figure}

Focussing on flavour charge assignments that allow for at least two (independent and non-trivial) flavour structures with up to four leptons, $d_\ell \leq 6$, see~\Cref{tbl:allowedOps}, \Cref{fig:rotary} illustrates the residual symmetries $G_\ell=\mathbb{Z}_N$ and flavour charge assignments compatible with these flavour structures. In doing so, we take into account the possibility of permutations among the charged lepton flavours.
 This figure highlights the restrictions that combinations of different observed cLFV processes could place. For example, it shows that if both tri-lepton decays $\tau \to \mu\mu \bar{e}$ and $\tau\to ee\bar{\mu}$ are observed, the only viable non-trivial flavour charge assignment is ${\bf 3(1,1)}$.\footnote{Also flavour charge assignments equivalent to ${\bf 3(1,1)}$ as  per~\cref{sec:notationTransformations} are viable (except that the permutation of the lepton flavours $e$, $\mu$ and $\tau$ is fixed). Concretely, for $G_\ell=\mathbb{Z}_3$ the flavour charge of each charged lepton cannot be uniquely determined, since the flavour charge assignment ${\bf 3(1,1)}$ corresponds to three different flavour charge combinations and, furthermore, ${\bf 3(2,2)}$ leads to the same allowed flavour structures, see Appendix~\ref{app:expandedNotation} and Section~\ref{sec:theory}. Thus, we can only conclude that the three charged leptons possess distinct flavour charges.} 
The table at the bottom of Figure~\ref{fig:rotary} lists the restrictions on the flavour charge assignments that result from a single flavour structure (corresponding to the observation of one type of cLFV processes). For instance, we clearly see that if at least one $\mu \to e$ transition is observed (the flavour structure $e\mu^\dagger$ is allowed), the muon and the electron have to have the same flavour charge, see also ${\bf 2(0,1)}$ in the figure, or no (approximate) residual symmetry $G_\ell$ is present among charged leptons. 

From~\Cref{fig:rotary}, one can see that in most instances, the observation of cLFV processes corresponding to only two different flavour structures is sufficient to constrain the size of the minimal equivalent residual symmetry to $\mathbb{Z}_N$ with $N=2$, $3$ or $4$, considering only lepton mass dimension $d_\ell\leq 6$. 
 All processes allowed by each of the flavour charge assignments for $G_\ell=\mathbb{Z}_4$ are also allowed by a corresponding flavour charge assignment for $G_\ell=\mathbb{Z}_2$. In order to distinguish these, the relevant radiative cLFV decays would have to be scrutinised.

Several combinations of  observed cLFV processes, corresponding to different types of flavour structures, could be used to (strongly) disfavour the existence of a  residual symmetry $G_\ell$\,---\,one that at least partially distinguishes between the three charged leptons. Arguably, the simplest such combination consists of two radiative cLFV decays.  Figure~\ref{fig:rotary} shows that cLFV processes corresponding to a maximum of five different flavour structures with $d_\ell \leq 6$
could be observed (which would be in the case of $G_\ell=\mathbb{Z}_2$) before excluding the existence of any (approximate) residual symmetries considered. 

The flavour charge assignments shown in Tables~\ref{tbl:allowedOps} and~\ref{tbl:allowedOperators} that do not explicitly appear in Figure~\ref{fig:rotary}, can be discussed according to the categories found in the table at the bottom of this figure.  Concretely, the flavour charge assignments $3(0,1)$, ${\bf 4(0,1)}$, ${\bf 5(0,1)}$, $6(0,1)$, ${\bf 7(0,1)}$ and ${\bf 8(0,1)}$ all only generate the flavour structure $e\mu^\dagger$ (take $a=1$ in the mentioned table), when restricting to dimension-six SMEFT operators. The analogous flavour structures, $\mu\tau^\dagger$ and $e\tau^\dagger$, are achieved by permuting the lepton flavours $e$, $\mu$ and $\tau$. 
 Consequently, the flavour structure $\mu\tau^\dagger$ is allowed by the flavour charge assignments $3(1,0)$, ${\bf 4(1,0)}$, ${\bf 5(1,0)}$, $6(1,0)$, ${\bf 7(1,0)}$ and ${\bf 8(1,0)}$, being of the form $N(a,0)$ with $a=1$. Likewise,
for the flavour structure $e \tau^\dagger$ viable flavour charge assignments are $3(1,2)$, ${\bf 4(1,3)}$, ${\bf 5(1,4)}$, $6(1,5)$, ${\bf 7(1,6)}$ and ${\bf 8(1,7)}$, meaning $N(a,N-a)$ for $a=1$ in agreement with the table at the bottom of Figure~\ref{fig:rotary}. The flavour structure $e\mu^\dagger\mu^\dagger\tau$ is allowed by the flavour charge assignments ${\bf 5(1,1)}$, ${\bf 6(1,1)}$, ${\bf 7(1,1)}$ and ${\bf 8(1,1)}$, all being of the form $N(a,a)$ with $a=1$. Applying permutations among the lepton flavours $e$, $\mu$ and $\tau$, the flavour structures $ee\mu^\dagger\tau^\dagger$ (up to Hermitian conjugation) and $e\mu\tau^\dagger\tau^\dagger$ arise.  Examples of flavour charge assignments allowing $ee\mu^\dagger\tau^\dagger$ as only (non-trivial) flavour structure with $d_\ell \leq 6$ are ${\bf 5(4,2)}$, ${\bf 6(5,2)}$, ${\bf 7(6,2)}$ and ${\bf 8(7,2)}$, whereas $e\mu\tau^\dagger\tau^\dagger$ is permitted by, for example, the flavour charge assignments ${\bf 5(2,4)}$, ${\bf 6(2,5)}$, ${\bf 7(2,6)}$ and ${\bf 8(2,7)}$. The two flavour charge assignments $6(1,2)$ and ${\bf 8(1,3)}$ both lead to the flavour structure $e e \tau^\dagger \tau^\dagger$, cf.~flavour charge assignment $2 N (a,N-a)$ with $a=1$ in the table at the bottom of Figure~\ref{fig:rotary}. Analogously, the two flavour structures $e e \mu^\dagger \mu^\dagger$ and $\mu \mu \tau^\dagger \tau^\dagger$, arising from permuting the lepton flavours $e$, $\mu$ and $\tau$, are allowed by $6(3,1)$ and ${\bf 8(4,1)}$ as well as $6(1,3)$ and ${\bf 8(1,4)}$, respectively, which follow the general forms $2N(N,a)$ and $2N(a,N)$, always with $a=1$, see the table at the bottom of Figure~\ref{fig:rotary}. 
 Lastly, we observe that the flavour charge assignments ${\bf 7(1,2)}$ and ${\bf 8(1,2)}$ do not allow for (non-trivial) flavour structures with $d_\ell \leq 6$. It is, thus, not possible to test these with the  cLFV processes discussed in this study.

We reiterate that while observations have the potential to constrain $G_\ell$ and the flavour charge assignments, these alone are not sufficient in order to arrive at any conclusion, since, at the same time, the cLFV processes corresponding to forbidden flavour structures have to be (strongly) suppressed.

\section{Sensitivity of cLFV searches to new physics scale}
\label{sec:boundsLambda}

In the preceding section, we have presented the inequivalent flavour charge assignments for $G_\ell=\mathbb{Z}_N$ that are compatible with flavour structures up to lepton mass dimension $d_\ell\leq6$.
 We now examine the bounds on the new physics scale $\Lambda$ that can be derived from various current and prospective cLFV experiments, see e.g.~\cite{Calibbi:2017uvl,Ardu:2022sbt,Davidson:2022jai} for reviews. In Section~\ref{lowenergycLFV} we focus on low-energy cLFV experiments, while we comment on possible cLFV searches at high-energy colliders in Section~\ref{sec:collider}.

Results for flavour charge assignments not explicitly mentioned in Figure~\ref{fig:rotary} can be found, for completeness, in Appendix~\ref{app:scale2}.
 
\subsection{Low-energy cLFV experiments}
\label{lowenergycLFV}

The tightest experimental limits on cLFV processes exist in the $\mu-e$ sector, with the strongest bound recently released by the MEG-II experiment on the decay $\mu^+\to e^+\gamma$, i.e.~${\rm BR}(\mu^+\to e^+\gamma) < 1.5 \times 10^{-13}~(90\% \,\text{C.L.})$~\cite{MEGII:2025gzr}, see~\Cref{app:scale}. This result is to be further improved by MEG-II itself~\cite{MEGII:2021fah}. 
The decay $\mu^+\to e^+e^+e^-$ is best constrained by measurements from SINDRUM, ${\rm BR}(\mu^+\to e^+e^+e^-) < 1.0 \times 10^{-12}~(90\% \,\text{C.L.})$~\cite{SINDRUM:1987nra}, and probes the same flavour transition as the decay $\mu^+\to e^+\gamma$, although $\mu^+\to e^+e^+e^-$ is currently less sensitive to the cLFV dipole operators. Upcoming searches by the Mu3e collaboration~\cite{Hesketh:2022wgw} are expected to improve this limit by three to four orders of magnitude, matching the prospective MEG-II sensitivity to the cLFV dipole operators, and largely exceeding MEG-II if other SMEFT operators are the dominant sources of cLFV.
Strong constraints in the $\mu-e$ sector are also provided by searches for $\mu-e$ conversion in nuclei $\rm N$, $\mu^-\, \mathrm{N} \to e^-\,\mathrm{N}$, with the strongest limit currently given by the SINDRUM-II experiment for $\mu-e$ conversion in gold, ${\rm CR}(\mu^-\,{\rm Au}\to e^-\,{\rm Au}) < 7 \times 10^{-13}~(90\% \,\text{C.L.})$~\cite{SINDRUMII:2006dvw}. This already stringent bound is foreseen to be improved by about four orders of magnitude by the Mu2e~\cite{Mu2e:2014fns} and COMET~\cite{COMET:2018auw} collaborations using aluminium targets.
Such an improvement corresponds to more than one order of magnitude increase in the sensitivity to the new physics scale $\Lambda$, see Appendix~\ref{app:scale}. In the long run, the Mu2e-II experiment can possibly achieve an even stronger limit by employing nuclei heavier than aluminium~\cite{Mu2e-II:2022blh}.
 A several orders of magnitude weaker constraint on the new physics scale $\Lambda$ in the $\mu-e$ sector is obtained from the bound on the muonium to antimuonium conversion probability, $\text{P}({\rm M}\rightarrow \overline{\rm M}) < 8.2\times 10^{-11}~(90\% \,\text{C.L.})$~\cite{Willmann:1998gd}. In the case in which $\mu\to e$ transitions are not allowed by the residual flavour symmetry $G_\ell$ and the flavour charge assignment of the charged leptons, the limit on $\text{P}({\rm M}\rightarrow \overline{\rm M})$ can be the 
 dominant constraint in the $\mu-e$ sector.
 Moreover, this bound is expected to be improved by about three orders of magnitude by the proposed MACE experiment~\cite{Bai:2024skk}.

The various cLFV $\tau$ lepton decays give rise to the next strongest constraints on the new physics scale $\Lambda$. These are, however, a few orders of magnitude weaker. 
The mass of the $\tau$ lepton allows for many possible decay channels, and thus experiments searching for them are vital in discriminating among residual flavour symmetries $G_\ell$ and flavour charge assignments 
 that forbid $\mu\to e$ transitions. The
current best limits on cLFV $\tau$ lepton decays, reaching the level $\order{10^{-8}}$, come from Belle, BaBar and Belle~II~\cite{BaBar:2006jhm,Belle:2007cio,Belle:2021ysv,Belle:2023ziz,BaBar:2009hkt,Belle:2010rxj,Hayasaka:2010np,Belle-II:2024sce}. All of these are expected to be improved by one to two orders of magnitude by future searches performed at Belle~II~\cite{Banerjee:2022xuw,Belle-II:2022cgf,Banerjee:2022vdd}. 

Other tests of cLFV can, in principle, be performed at high-energy collider experiments searching for cLFV scattering processes and cLFV decays of heavy SM bosons. Current constraints from these sources, where they exist, are significantly weaker than those from muon and $\tau$ lepton decays, see Section~\ref{sec:collider}.

\begin{figure}[t!]
   \centering
   \includegraphics[width=\textwidth]{"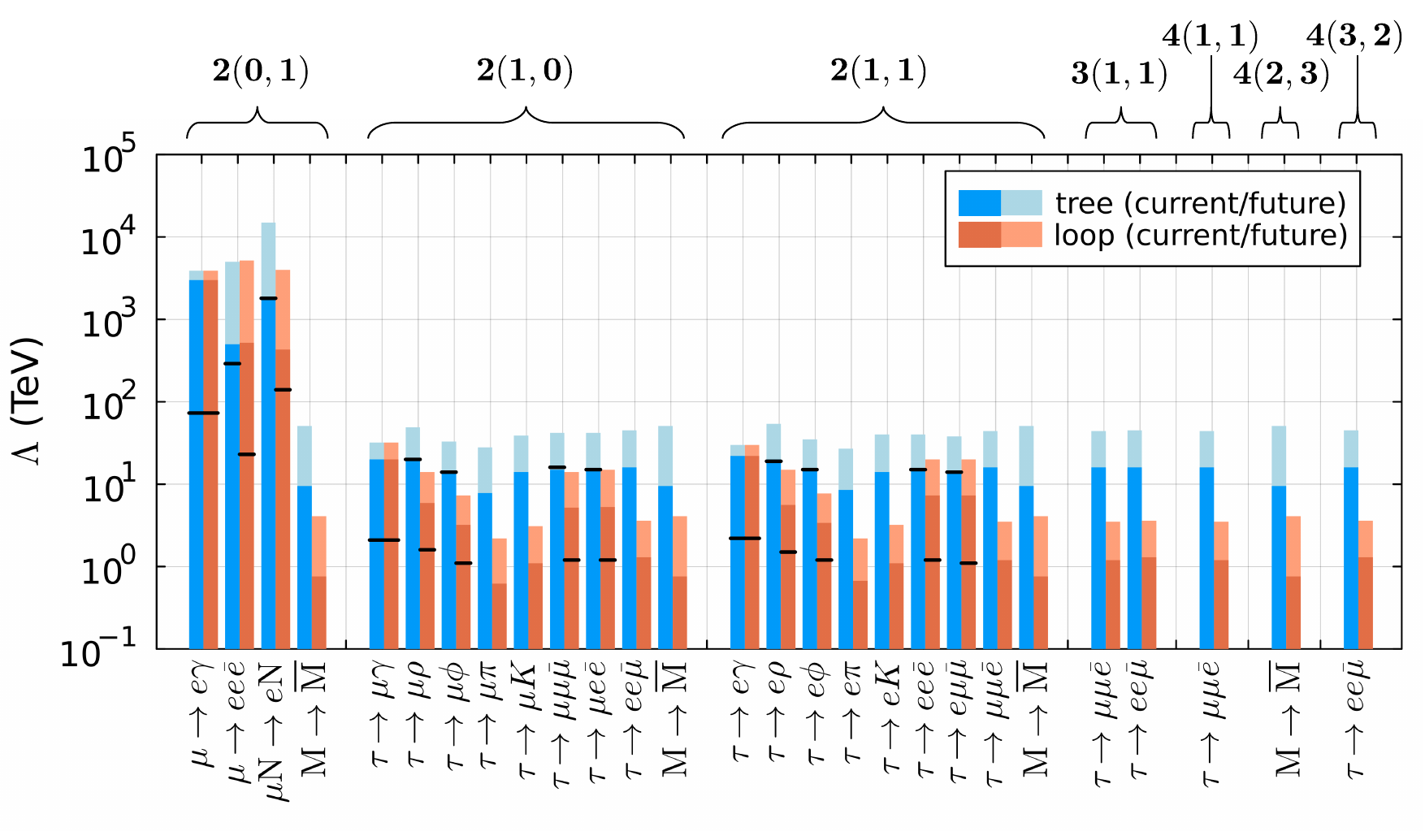"}
   \caption{Current 
    lower bounds on and future sensitivities to the new physics scale $\Lambda$ (in~$\rm TeV$), based on experimental limits for various processes at $90\%\,\rm C.L.$.
   The results are grouped according to the residual flavour symmetry $G_\ell$ and flavour charge assignment of the charged leptons, cf.~\Cref{tbl:allowedOps} and~\Cref{fig:rotary}.
 Present bounds (dark colours) and future sensitivities (light colours) are shown for the two different scenarios with tree-level (blue) and loop-level contributions (orange). The results for the scenario with chirally suppressed contributions are indicated by a horizontal black line, where relevant. See  text for details and~\Cref{app:scale} for the numerical values of the new physics scale $\Lambda$ in the different considered scenarios.}
   \label{fig:scalebar}
\end{figure}

For each of the mentioned processes, we can calculate the corresponding observable in terms of the WCs of the contributing SMEFT operators and the new physics scale $\Lambda$, cf.~\cref{eq:smeft}, by means of the formulae collected in~\cite{Calibbi:2021pyh}. We then translate the experimental upper bound on the observable into a lower bound on $\Lambda$, making different well-motivated assumptions about the WCs. The results are shown in~\Cref{fig:scalebar}.

We primarily consider two scenarios in which new physics contributes at the tree and the one-loop level,  respectively. In the case of tree-level new physics, we set all WCs of the operators allowed by a certain residual symmetry $G_\ell$ and flavour charge assignment 
 to $C_x=1$, except for those of the dipole operators that are generated at (one-)loop level within any realistic UV-complete model. Hence, we set $C_{d}=e/(16\pi^2)\approx 0.002$, where $e=\sqrt{4\pi\alpha} \approx 0.30$ is the electromagnetic coupling at low energy. In the case that all new physics effects arise at the one-loop level, the WCs of the dipole operators remain the same, while all other (non-zero) WCs become suppressed by a loop factor, $C_x=1/(16\pi^2) \approx 0.006$.

It is worth also considering a third scenario, in which the WCs of the dipole operators are further suppressed by the Yukawa coupling of the decaying lepton, representing a mass insertion as required to perform a chirality flip. This leads to $C_{d} = \sqrt{2} \, m_\ell \, e/(16\pi^2 v)$. The effects of this further suppression are shown by black lines in~\Cref{fig:scalebar} for the processes with contributions from the dipole operators. This type of suppression is negligible, if the WCs of the non-dipole operators are already dominant, as it is the case for several $\tau$ lepton decays. 

Finally, note that we neither consider  renormalisation group (RG) running of the WCs from the new physics scale $\Lambda$ to the relevant energy scale of a given process nor the matching of SMEFT operators to Low-energy Effective Field Theory operators at the electroweak scale. Since in the current approach all operators allowed by 
 a certain residual symmetry $G_\ell$ and flavour charge assignment are present in the UV with a comparable strength and, furthermore, $G_\ell$
  prohibits forbidden operators to arise through RG running, this simplifying assumption only makes us overlook corrections to the WCs e.g.~of the order $\alpha_{(s)}/(4\pi)$~\cite{Alonso:2013hga}. These are numerically negligible, even when stemming from perturbative QCD.

In~\Cref{fig:scalebar}, we display the lower bound on and future sensitivity to the new physics scale $\Lambda$, obtained from various relevant cLFV searches. Comparing the values of $\Lambda$ across the three different scenarios (with tree-level, loop-level or chirally suppressed contributions) provides an idea of the model dependence of the results. The (current/prospective) experimental bounds and numerical values for $\Lambda$ are tabulated in~\Cref{app:scale}. 

Following the discussion in~\Cref{sec:classifyops}, the processes in~\Cref{fig:scalebar} are grouped according to the residual flavour symmetry $G_\ell=\mathbb{Z}_N$ and the flavour charge assignments they are compatible with. In this way, we can contrast the resulting selection rules with the reach of present and future cLFV experiments.

Since in all considered cases the present experimental limits entail $\Lambda$ larger than the electroweak scale, the adoption of SMEFT
is justified. As expected, the limits on $\Lambda$ arising from $\mu\to e$ transitions are far stronger than those of other processes, predominately $\tau$ lepton decays, which is consistent with similar analyses available in the literature, see e.g.~\cite{Carpentier:2010ue,Crivellin:2013hpa,Pruna:2014asa,Feruglio:2015gka,Crivellin:2017rmk,Davidson:2018rqt,Ardu:2021koz,Ardu:2022pzk,Jahedi:2024kvi,Calibbi:2021pyh,Calibbi:2022ddo}.\footnote{Nevertheless, numerical differences are expected with respect to these studies because of the different treatment of the WCs of the dipole operators and, more importantly, the choice of including all allowed SMEFT operators, whereas previous studies typically display results obtained by considering a single  operator at a time.}
 However, we see that $\mu\to e$ transitions are only allowed by flavour charge assignments such as ${\bf 2(0,1)}$ that forbid cLFV $\tau$ lepton decays. On the contrary, flavour charge assignments such as ${\bf 2(1,0)}$ can be best probed with $\tau$ lepton decays, showing that cLFV searches for processes involving $\tau$ leptons can be the key in order to underpin the existence of a certain residual symmetry $G_\ell$ and flavour charge assignment. In particular, Figure~\ref{fig:scalebar} shows that if one cLFV $\tau$ lepton decay is observed at Belle~II, while not observing several more, this would disfavour $G_\ell=\mathbb{Z}_2$ as residual symmetry among charged leptons.
  Moreover, cLFV searches for the tri-lepton decays $\tau\to ee\bar \mu$ and $\tau\to \mu\mu\bar e$ are suitable to test flavour charge assignments associated with the residual symmetries $G_\ell=\mathbb{Z}_3$ and $G_\ell=\mathbb{Z}_4$.
In general, along with stronger constraints on muonium to antimuonium conversion, stronger constraints on $\tau$ lepton decays could provide information on the plausibility of all the discussed flavour charge assignments. Furthermore, 
 we note that, although several residual symmetries $G_\ell$ and flavour charge assignments allow muonium to antimuonium conversion, this quantitative study shows that stringent bounds
 from $\mu\to e$ transitions render this process unobservable for the flavour charge assignment ${\bf 2(0,1)}$ and $G_\ell=\mathbb{Z}_2$, while it could be observed, but alongside cLFV $\tau$ lepton decays, if
 either the flavour charge assignment ${\bf 2(1,0)}$ or ${\bf 2(1,1)}$ is (approximately) realised in Nature.
 Additionally, experimental evidence of muonium to antimuonium conversion alone is compatible with (and could provide support for) the flavour charge assignment ${\bf 4(2,3)}$ and $G_\ell=\mathbb{Z}_4$.

\subsection{High-energy colliders}
\label{sec:collider}

The limits in~\Cref{fig:scalebar} do not take into account SMEFT operators corresponding to flavour structures in~\Cref{tbl:allowedOps} that contain two \emph{same sign} $\tau$ leptons. These would be directly tested, when searching for the scattering processes
\begin{align}
    e^\pm e^\pm \to \tau^\pm \tau^\pm\,,
    \quad e^\pm \mu^\pm \to \tau^\pm \tau^\pm\,,
    \quad \mu^\pm \mu^\pm \to \tau^\pm \tau^\pm\,.
    \label{eq:scatt}
\end{align}
There are no current colliders at  which any of these could be probed. One may argue that tests of the SMEFT operators involving two $\tau$ leptons are not strictly necessary. As mentioned, 
 the flavour charge assignments belonging to $G_\ell=\mathbb{Z}_4$
 may be distinguished from the corresponding ones for $G_\ell=\mathbb{Z}_2$ by scrutinising the relevant radiative cLFV decays. However, the preceding discussion has also shown that the 
 observation of cLFV processes corresponding to distinct flavour structures is necessary to build up potential evidence of a certain residual symmetry $G_\ell$ and flavour charge assignment.
 If observed in combination with, respectively, $\tau \to\mu \mu \bar e$,  muonium to antimuonium conversion or $\tau \to ee\bar \mu$, the three processes in Eq.~(\ref{eq:scatt}) would be a solid indication for one of the flavour charge assignments for $G_\ell=\mathbb{Z}_4$, as illustrated by~\Cref{fig:rotary}.

An environment in which part of the processes shown in Eq.~(\ref{eq:scatt}) could possibly be searched for is the proposed $\mu$TRISTAN project~\cite{Hamada:2022mua}. The proposal involves two operation modes: 
(i)~employing two 1~TeV $\mu^+$ beams to deliver $\mu^+\mu^+$ collisions with centre-of-mass (c.o.m.)~energy $\sqrt{s} = 2$~TeV; (ii)~colliding a 1~TeV $\mu^+$ beam with a 30~GeV $e^-$ beam achieving a Higgs factory with $\sqrt{s} \simeq 346$~GeV.
The first operation mode could directly be used to search for $\mu^+\mu^+ \to \tau^+\tau^+$ and $\mu^+\mu^+ \to e^+e^+$. The flavour structure corresponding to the latter process
is, however, already probed by muonium to antimuonium  conversion. A modification of the second operation mode employing an $e^+$ beam instead of an $e^-$ beam could, in principle, be used to constrain the process $e^+\mu^+ \to \tau^+\tau^+$. When induced, as in the case at hand, by four-lepton SMEFT operators, e.g.~$(\bar{\ell}_\tau\gamma_\nu\ell_\mu)(\bar{\ell}_\tau\gamma^\nu\ell_\mu)$, these scattering cross sections increase with the c.o.m.~energy squared\footnote{Notice that this is consistent with unitarity as long as $\sqrt{s} \ll \Lambda$, that is, in the regime of validity of SMEFT. For a discussion of processes induced by cLFV SMEFT operators in the context of high-energy $e^+e^-$ colliders, we refer to~\cite{Altmannshofer:2023tsa}. See also~\cite{Lichtenstein:2023iut,Fridell:2023gjx} for studies on possible searches for cLFV at $\mu$TRISTAN.}
\begin{align}
    \sigma(\mu^+\mu^+ \to \tau^+\tau^+) = \frac{s}{2\pi} \frac{|C_x|^2}{\Lambda^4} \simeq 25\,\text{fb} \,\left(\frac{\sqrt{s}}{2\,\text{TeV}}\right)^2\left(\frac{10\,\text{TeV}}{\Lambda/\sqrt{|C_x|}}\right)^4 \,,
\end{align}
while for $\sigma(e^+\mu^+ \to \tau^+\tau^+)$ this formula should be divided by a factor of four, because the initial-state particles are not identical. 
From here, we can see that $\mu$TRISTAN would have the capability to test the scattering $\mu^+\mu^+ \to \tau^+\tau^+$ up to $\Lambda\approx 30$~TeV for $C_x =1$.\footnote{Since the signature is free of (irreducible) physics backgrounds, we conservatively assume a selection/detection efficiency of $\varepsilon_\text{eff} = 10\%$, and that evidence for the process can be achieved for a number of signal events \mbox{$n_s = \sigma(\mu^+\mu^+ \to \tau^+\tau^+)\times\varepsilon_\text{eff}\times \mathcal{L} = 3$}, where the expected integrated luminosity is $\mathcal{L}=100\,\text{fb}^{-1}$~\cite{Hamada:2022mua}.} Hence, it probes scales comparable to the expected sensitivities of searches for cLFV $\tau$ lepton decays and muon to antimuonium conversion.

Alternatively, SMEFT operators with flavour structures such as $ee\tau^{\dag}\tau^{\dag}$ and $\mu\mu\tau^{\dag}\tau^{\dag}$ could be tested in heavy boson decays. The example of the four-body $Z$ decay $Z\to \tau\tau\bar{e}\bar{e}$, and the analogous decay $Z\to \tau\tau\bar{\mu}\bar{\mu}$, is discussed in~\cite{Heeck:2024uiz}. However, currently there are no experimental limits on these processes, and requiring them not to be in conflict with the well-measured (and well-predicted) $Z$ decay width only leads to the extremely weak constraint $\Lambda \gtrsim 1$~GeV.
If future Tera-Z factories, such as CEPC~\cite{CEPCStudyGroup:2023quu} and FCC-ee~\cite{FCC:2018byv}, can set a limit $\text{BR}(Z\to \tau\tau\bar{e}\bar{e}) < 10^{-12}$, the constraint would be strongly improved, $\Lambda \gtrsim 250$~GeV~\cite{Heeck:2024uiz}. Still, such a sensitivity seems to be insufficient to exceed the current limits from $\tau$ lepton decays, shown in~\Cref{fig:scalebar}, and is just barely 
within the regime of validity of SMEFT.

\section{Summary and outlook}
\label{sec:summ}

Predictive models of lepton flavour with a discrete flavour (and CP) symmetry often feature a residual symmetry $G_\ell$ for charged leptons, which restricts the allowed cLFV processes. We have analysed the resulting selection rules and phenomenological consequences for
all possible flavour charge assignments and $G_\ell=\mathbb{Z}_N$ with $N\leq 8$. These reduce to a small set of unique flavour charge assignments. For each, we have provided a complete categorisation of the allowed flavour structures, compare Tables~\ref{tbl:allowedOps} and~\ref{tbl:allowedOperators}. In the phenomenological analysis, we have matched each flavour structure with no more than four charged leptons to SMEFT operators, up to dimension six, see~\Cref{fig:rotary}. We have found that the observation of cLFV processes corresponding to more than one of these flavour structures can constrain any combination of flavour charge assignment and (approximate) residual symmetry $G_\ell=\mathbb{Z}_N$ to be equivalent to one with $N\leq 4$.
 The results illustrate the patterns of cLFV processes that, if observed, could point to the existence of a certain
residual symmetry $G_\ell$ and flavour charge assignment for the charged leptons. 

The quantitative analysis in Section~\ref{sec:boundsLambda} has  demonstrated that evidence for the obtained selection rules can indeed be achieved with current and near-future experiments.
The lower limits on and prospective sensitivities to the new physics scale $\Lambda$ shown in~\Cref{fig:scalebar} can be interpreted as the scale of a possible UV completion of the lepton flavour theory. At the same time, they indicate 
 the order of magnitude of the breaking scale of the employed flavour (and CP) symmetry. An important observation is that flavour charge assignments which distinguish among the electron and the muon do not allow $\mu\to e$ transitions. As a consequence, searches for cLFV $\tau$ lepton processes and muonium to antimuonium conversion can be crucial to provide complementary information. 

Let us comment on directions in which this study is worth extending. On the one hand, one could consider concrete models with $G_\ell$ being one of the analysed symmetries, and possibly focus on specific UV completions. On the other hand, one could study effective operators that also violate lepton number (in general, assuming a different scale $\Lambda_{\mathrm{LNV}}$) since these may have interesting phenomenological consequences. For example, neutrinoless double beta decay could be forbidden and $\mu^+$ to $e^-$ conversion in nuclei allowed, such that evidence for lepton number violation could be first observed at conversion experiments such as Mu2e~\cite{Mu2e:2014fns} and COMET~\cite{COMET:2018auw}. Furthermore, the consideration of lepton number violating processes may distinguish flavour charge assignments of the charged leptons that lead to the same set of allowed lepton number conserving processes. Additionally, one could apply a similar logic to the quark sector and discuss processes of quark flavour violation. In doing so, one possibility is to apply the residual symmetry of the charged leptons, and potentially also their flavour charge assignment, to quarks.\footnote{Consequently, quark mixing would vanish at the level of exact residual symmetries, see e.g.~\cite{Feruglio:2007uu}, and a small breaking could explain the smallness of the quark mixing angles.} Another option is to have different residual symmetries for up and down quarks, as for example in~\cite{Blum:2007jz,Hagedorn:2012pg}, which generally also differ from the symmetry $G_\ell$. 

\vspace{0.1in}
\paragraph{Acknowledgements.} L.C.~and C.H.~thank the School of Physics of the University of New South Wales (UNSW) for its hospitality and financial support during the initial stage of this project.
L.C.~acknowledges financial support
from the National Natural Science Foundation of China (NSFC) under the grant No.~12035008.
C.H.~is supported by the Spanish MINECO through the Ram\'o{}n y Cajal programme RYC2018-024529-I, by the national grant PID2023-148162NB-C21, by the Generalitat Valenciana through PROMETEO/2021/083, by the MCIU/AEI Severo-Ochoa-project CEX2023-001292-S as well as by the European Union's Horizon 2020 research and innovation programme under the Marie Sk\l{}odowska-Curie grant agreement No.~860881 (HIDDe$\nu$ network) and under the Marie Sk\l{}odowska-Curie Staff Exchange grant agreement No.~101086085 (ASYMMETRY). M.S.~acknowledges support by the Australian Research Council Discovery Project DP200101470. 
J.V.~is supported by an Australian Government Research Training Program (RTP) Scholarship.


\appendix

\section{Completeness of study}
\label{app:flavProof}

 Here, we show that permitted flavour structures with at least one $|\Delta n_{e,\mu,\tau}| > N$ can be decomposed into a combination of allowed flavour structures with $|\Delta n_{e,\mu,\tau}|\leq N$. Consider an arbitrary flavour charge assignment $\mathbb{Z}_N(\alpha,\beta,\gamma)$ and an allowed flavour structure encoded in $ \{\Delta n_e',\Delta n_\mu', \Delta n_\tau'\}$. Since this flavour structure is allowed,  the constraints in~\cref{eq:echargeconstraint,eq:fchargeconstraint} are fulfilled, i.e.~ 
\begin{equation}
 \Delta n_e' + \Delta n_\mu' + \Delta n_\tau' = 0 \;\;  \mbox{and} \;\; \alpha \, \Delta n_e' + \beta \, \Delta n_\mu' + \gamma \, \Delta n_\tau' = 0 \text{ mod } N\ .
\label{eq:appCfconstraint}
\end{equation}
 For integers $a$, $b$, $c$, we can write the changes in the individual charged lepton numbers as
\begin{equation}
\Delta n_e' = a \, N + \Delta n_e\, , \;\, \Delta n_\mu' = b \, N + \Delta n_\mu \, , \;\, \Delta n_\tau' = c \, N + \Delta n_\tau \ , 
\label{eq:reparam}
\end{equation}
where $0 \leq \Delta n_e \leq N$ and $-N \leq \Delta n_\mu, \, \Delta n_\tau \leq N$.
We have to show that integers $a$, $b$, $c$ and $\Delta n_e$, $\Delta n_\mu$, $\Delta n_\tau$ exist such that the latter also correspond to an allowed flavour structure, i.e.~
\begin{align}
\Delta n_e + \Delta n_\mu + \Delta n_\tau = 0 \ ,\label{eq:appCeconstraint2}\\
\alpha \, \Delta n_e + \beta \, \Delta n_\mu + \gamma \, \Delta n_\tau &= 0 \text{ mod }N \; .\label{eq:appCfconstraint2}
\end{align}
Plugging~\cref{eq:reparam} into the second equation in~\cref{eq:appCfconstraint}, we immediately find
\begin{align}
    \alpha \, \Delta n_e + \beta \, \Delta n_\mu + \gamma \, \Delta n_\tau &= 0 \text{ mod }N \ ,
\end{align}
meaning that $\Delta n_{e,\mu,\tau}$ conserve the flavour charge (independent of the values of $a$, $b$ and $c$). Plugging~\cref{eq:reparam} into the first equation in~\cref{eq:appCfconstraint}, we instead get
\begin{align}
(a+b+c) + \frac{\Delta n_e + \Delta n_\mu + \Delta n_\tau}{N} &= 0 \; . \label{eq:asandns}
\end{align}
We discuss the implications of~\cref{eq:asandns} on a case by case basis. First, note that if 
\begin{equation}
\label{eq:DeltanasnotzN}
\Delta n_e + \Delta n_\mu + \Delta n_\tau \neq z \, N
\end{equation}
with $z$ being an integer, it must hold that $a+b+c = 0$ and $\Delta n_e + \Delta n_\mu + \Delta n_\tau = 0$ separately, as the former is integer-valued, and the latter, when divided by $N$, is not. 
If so, also $\Delta n_{e,\mu,\tau}$ fulfil the constraint on the charged lepton number, see Eq.~(\ref{eq:appCeconstraint2}). 
On the other hand, if 
\begin{equation}
\label{eq:DeltanaszN}
\Delta n_e + \Delta n_\mu + \Delta n_\tau = z \, N \, , 
\end{equation}
then $a+b+c=-z$ must be true and by definition of $\Delta n_{e,\mu,\tau}$, the integer $z$ can only take values in the interval $-2\leq z \leq 3$.
For $z=0$, $\Delta n_{e,\mu,\tau}$ satisfy the condition on the charged lepton number immediately.
If $z=1 \, (-1)$, we can assume that $\Delta n_\mu \geq 0$ ($\Delta n_\mu \leq 0$). We can then reparameterise $\Delta n_\mu$ such that $\Delta n_\mu \to \Delta n_\mu - N$ and $b\to b+1$  ($\Delta n_\mu \to \Delta n_\mu + N$ and $b\to b-1$). This reparameterisation leads back to the case of $z=0$. 
For $z=2 \, (-2)$, both $\Delta n_\mu$ and $\Delta n_\tau$ must be positive (negative). In a similar fashion to the previous case, we can modify $\Delta n_{\mu, \tau}$, i.e.~$\Delta n_\mu \to \Delta n_\mu - N$ and $\Delta n_\tau \to \Delta n_\tau - N$ together with $b\to b+1$ and  $c\to c+1$ ($\Delta n_\mu \to \Delta n_\mu + N$ and $\Delta n_\tau \to \Delta n_\tau + N$ together with $b\to b-1$ and $c\to c-1$), and apply again the results for $z=0$. 
In the case where $\Delta n_e = N \, (0)$, we could also remove (add) $N$ from (to) $\Delta n_e$ and one of either $\Delta n_\mu$ or $\Delta n_\tau$. 
Finally, for $z=3$, necessarily $\Delta n_{e,\mu,\tau} = N$. 
Then, we may choose to reparameterise $\Delta n_{e,\mu,\tau}$ by removing either $N$ from each $\Delta n_{e,\mu,\tau}$ such that $\Delta n_{e,\mu,\tau} = 0$, or we may remove $N$ from $\Delta n_e$ and $2 \, N$ from either $\Delta n_\mu$ or $\Delta n_\tau$, or remove $N$ from either $\Delta n_\mu$ or $\Delta n_\tau$ and $2\, N$ from the other one, while not changing $\Delta n_e$. All of these reparametrisations restore the situation of $z=0$.
    
\newpage

\section{\mathversion{bold} Allowed flavour structures for \texorpdfstring{$G_\ell=\mathbb{Z}_{6,7,8}$}{GlZ678} \texorpdfstring{(extension of~\cref{tbl:allowedOps})}{extension}\mathversion{normal}}
\label{app:allowedOps}

\begin{table}[htbp!]
\renewcommand\arraystretch{0.85}
\setlength{\tabcolsep}{20pt}
        \begin{tabular}{cc}
\setlength{\tabcolsep}{6pt}    
\begin{tabular}{lcl}
        \toprule
                Flavour & & Flavour\\
        charges & $d_\ell$ & structures\\\midrule
\multirow[t]{8}{8em}{$6(0, 1)$}
 & $3$ &$e\mu^\dagger $\\
 & $18$ &$\mu^6(\tau^\dagger)^6 $\\
 &  &$e\mu^5(\tau^\dagger)^6 $\\
 &  &$e^2\mu^4(\tau^\dagger)^6$\\
 &  &$e^3\mu^3(\tau^\dagger)^6$\\
 &  &$e^4\mu^2(\tau^\dagger)^6$\\
 &  &$e^5\mu(\tau^\dagger)^6$\\
 &  &$e^6(\tau^\dagger)^6$\\
\midrule
\multirow[t]{8}{8em}{\textbf{$\mathbf{6( 1, 1)}$}}
 & $6$ &$e(\mu^\dagger)^2\tau$\\
 & $9$ &$e^3(\tau^\dagger)^3 $\\
 & $12$ &$e^2\mu^2(\tau^\dagger)^4$\\
 &  &$e^4(\mu^\dagger)^2(\tau^\dagger)^2$\\
 & $15$ &$e\mu^4(\tau^\dagger)^5$\\
 &  &$e^5(\mu^\dagger)^4\tau^\dagger $\\
 & $18$ &$\mu^6(\tau^\dagger)^6$\\
 &  &$e^6(\mu^\dagger)^6$\\
\midrule
\multirow[t]{5}{8em}{$6( 1, 2)$}
 & $6$ &$e^2(\tau^\dagger)^2$\\
 & $9$ &$\mu^3(\tau^\dagger)^3 $\\
 &  &$e^2(\mu^\dagger)^3\tau$\\
 & $12$ &$e^4(\mu^\dagger)^3 \tau^\dagger $\\
 & $18$ &$e^6(\mu^\dagger)^6$\\
\midrule
\multirow[t]{9}{8em}{$\mathbf{7( 0, 1)}$}
 & $3$ &$e\mu^\dagger $\\
 & $21$ &$\mu^7(\tau^\dagger)^7$\\
 &  &$e\mu^6(\tau^\dagger)^7$\\
 &  &$e^2\mu^5(\tau^\dagger)^7$\\
 &  &$e^3\mu^4(\tau^\dagger)^7$\\
 &  &$e^4\mu^3(\tau^\dagger)^7$\\
 &  &$e^5\mu^2(\tau^\dagger)^7$\\
 &  &$e^6\mu(\tau^\dagger)^7$\\
 &  &$e^7(\tau^\dagger)^7$\\
\midrule
\multirow[t]{10}{8em}{$\mathbf{7( 1, 1)}$}
 & $6$ &$e(\mu^\dagger)^2\tau$\\
 & $12$ &$e^3\mu(\tau^\dagger)^4$\\
 &  &$e^4\mu^\dagger (\tau^\dagger)^3$\\
 & $15$ &$e^2\mu^3(\tau^\dagger)^5$\\
 &  &$e^5(\mu^\dagger)^3 (\tau^\dagger)^2$\\
 & $18$ &$e\mu^5(\tau^\dagger)^6$\\
 &  &$e^6(\mu^\dagger)^5 \tau^\dagger$\\
 & $21$ &$\mu^7(\tau^\dagger)^7$\\
 &  &$e^7(\tau^\dagger)^7$\\
 &  &$e^7(\mu^\dagger)^7 $\\
\bottomrule
\end{tabular}&
\setlength{\tabcolsep}{6pt}
\begin{tabular}{lcl}
        \toprule
                Flavour & & Flavour\\
        charges & $d_\ell$ & structures\\\midrule
\multirow[t]{9}{8em}{$\mathbf{7( 1, 2)}$}
 & $9$ &$e\mu^2(\tau^\dagger)^3$\\
 &  &$e^2(\mu^\dagger)^3 \tau$\\
 &  &$e^3\mu^\dagger (\tau^\dagger)^2 $\\
 & $15$ &$e(\mu^\dagger)^5 \tau^4$\\
 &  &$e^4\mu(\tau^\dagger)^5$\\
 &  &$e^5(\mu^\dagger)^4\tau^\dagger $\\
 & $21$ &$\mu^7(\tau^\dagger)^7$\\
 &  &$e^7(\tau^\dagger)^7$\\
 &  &$e^7(\mu^\dagger)^7$\\
\midrule
\multirow[t]{10}{8em}{$\mathbf{8( 0, 1)}$}
 & $3$ &$e\mu^\dagger $\\
 & $24$ &{\small$\mu^8(\tau^\dagger)^8$}\\
 &  &{\small$e\mu^7(\tau^\dagger)^8$}\\
 &  &{\small$e^2\mu^6(\tau^\dagger)^8$}\\
 &  &{\small$e^3\mu^5(\tau^\dagger)^8$}\\
 &  &{\small$e^4\mu^4(\tau^\dagger)^8$}\\
 &  &{\small$e^5\mu^3(\tau^\dagger)^8$}\\
 &  &{\small$e^6\mu^2(\tau^\dagger)^8$}\\
 &  &{\small$e^7\mu(\tau^\dagger)^8$}\\
 &  &{\small$e^8(\tau^\dagger)^8$}\\
\midrule
\multirow[t]{10}{8em}{$\mathbf{8( 1, 1)}$}
 & $6$ &$e(\mu^\dagger)^2\tau$\\
 & $12$ &$e^4(\tau^\dagger)^4$\\
 & $15$ &$e^3\mu^2(\tau^\dagger)^5$\\
 &  &$e^5(\mu^\dagger)^2 (\tau^\dagger)^3$\\
 & $18$ &$e^2\mu^4(\tau^\dagger)^6$\\
 &  &$e^6(\mu^\dagger)^4(\tau^\dagger)^2$\\
 & $21$ &$e\mu^6(\tau^\dagger)^7$\\
 &  &$e^7(\mu^\dagger)^6 \tau^\dagger $\\
 & $24$ &$\mu^8(\tau^\dagger)^8$\\
 &  &$e^8(\mu^\dagger)^8$\\
\midrule
\multirow[t]{8}{8em}{$\mathbf{8(1, 2)}$}
 & $9$ &$e^2\mu(\tau^\dagger)^3$\\
 &  &$e^2(\mu^\dagger)^3\tau$\\
 & $12$ &$\mu^4(\tau^\dagger)^4$\\
 &  &$e^4(\mu^\dagger)^2 (\tau^\dagger)^2 $\\
 & $18$ &$e^6\mu^\dagger (\tau^\dagger)^5 $\\
 &  &$e^6(\mu^\dagger)^5 \tau^\dagger $\\
 & $24$ &$e^8(\tau^\dagger)^8$\\
 &  &$e^8(\mu^\dagger)^8$\\
\midrule
\multirow[t]{7}{8em}{$\mathbf{8( 1, 3)}$}
 & $6$ &$e^2(\tau^\dagger)^2 $\\
 & $12$ &$e(\mu^\dagger)^4 \tau^3$\\
 &  &$e^3(\mu^\dagger)^4 \tau$\\
 & $15$ &$e\mu^4(\tau^\dagger)^5$\\
 &  &$e^5(\mu^\dagger)^4 \tau^\dagger $\\
 & $24$ &$\mu^8(\tau^\dagger)^8$\\
 &  & $e^8(\mu^\dagger)^8$\\
\bottomrule
\end{tabular}
\end{tabular}
        \caption{Allowed flavour structures for $G_\ell=\mathbb{Z}_N$,  $6\leq N\leq 8$. We use a shorthand notation for repeated lepton flavours, e.g.~$e^2=ee$. For further details, see~\Cref{tbl:allowedOps} and main text.}
        \label{tbl:allowedOperators}
        \centering
            
\end{table}

\clearpage

\section{\mathversion{bold}Flavour charge assignments for \texorpdfstring{$G_\ell=\mathbb{Z}_3$}{GlZ3}\mathversion{normal}}
\label{app:expandedNotation}

\begin{table}[ht!] 
       \centering
       \renewcommand\arraystretch{1}
       \begin{tabular}{ccc}
               \toprule
               $P_1$ & $P_2$ & $P_3$ \\\midrule
           \begin{tabular}{lc}
$N(\delta_{1},\delta_{2})$ & Explicit form\\\midrule
$3(0,1)$ & $\mathbb{Z}_3(0,0,1)$\\
                               & $\mathbb{Z}_3(1,1,2)$\\
                               & $\mathbb{Z}_3(2,2,0)$\vspace{0.1cm}\\
$3(0,2)$ & $\mathbb{Z}_3(0,0,2)$\\
                               & $\mathbb{Z}_3(1,1,0)$\\
                               & $\mathbb{Z}_3(2,2,1)$
           \end{tabular} & 
           \begin{tabular}{lc}
$N(\delta_{1},\delta_{2})$ & Explicit form\\\midrule
$3(1,2)$ & $\mathbb{Z}_3(0,1,0)$\\
                               & $\mathbb{Z}_3(1,2,1)$\\
                            & $\mathbb{Z}_3(2,0,2)$\vspace{0.1cm}\\
$3(2,1)$ & $\mathbb{Z}_3(0,2,0)$\\
                               & $\mathbb{Z}_3(1,0,1)$\\
                               & $\mathbb{Z}_3(2,1,2)$\vspace{0.1cm}\\
           \end{tabular} & 
           \begin{tabular}{lc}
$N(\delta_{1},\delta_{2})$ & Explicit form\\\midrule
$3(2,0)$ & $\mathbb{Z}_3(1,0,0)$\\
                               & $\mathbb{Z}_3(2,1,1)$\\
                        & $\mathbb{Z}_3(0,2,2)$\vspace{0.1cm}\\
$3(1,0)$ & $\mathbb{Z}_3(0,1,1)$\\
                               & $\mathbb{Z}_3(1,2,2)$\\
                        & $\mathbb{Z}_3(2,0,0)$\vspace{0.1cm}\\
\end{tabular}\vspace{0.1cm}\\
               \toprule
                  & $P_{1,2,3}$ & \\\midrule
&
           \begin{tabular}{lc}
$N(\delta_{1},\delta_{2})$ & Explicit form\\\midrule
$\mathbf{3(1,1)}$ & $\mathbf{\mathbb{Z}_3(0,1,2)}$\\
                               & $\mathbf{\mathbb{Z}_3(1,2,0)}$\\
                            & $\mathbf{\mathbb{Z}_3(2,0,1)}$\vspace{0.1cm}\\
$\mathbf{3(2,2)}$ & $\mathbf{\mathbb{Z}_3(0,2,1)}$\\
                               & $\mathbf{\mathbb{Z}_3(1,0,2)}$\\
                               & $\mathbf{\mathbb{Z}_3(2,1,0)}$\vspace{0.1cm}\\
           \end{tabular}
 &
           \\\bottomrule
\end{tabular}
\caption{All flavour charge assignments for $G_\ell=\mathbb{Z}_3$. $P_1$, $P_2$, and $P_3$ label the permutations of the charged lepton flavours.}
\label{tbl:ExpandedGroups3}        
\end{table}

\clearpage

\section{Numerical values for \mathversion{bold}\texorpdfstring{$\Lambda$}{Lambda}\mathversion{normal}
from current and future cLFV searches}
\label{app:scale}

In~\cref{sec:boundsLambda}, the calculation of lower limits on and future sensitivities to the new physics scale $\Lambda$ is discussed, and the results are displayed in Figure~\ref{fig:scalebar}.
In this appendix, the experimental limits employed and the numerical values of $\Lambda$ are tabulated in~\cref{tbl:scaleData,tbl:scaleDataCollider}.

\begin{table}[ht!]
   \centering
   \renewcommand\arraystretch{1.}
        \setlength{\tabcolsep}{2pt}
   \resizebox{\linewidth}{!}{\begin{tabular}{crlcccccc}
   \toprule
     \multirow{2}{*}{$N(\delta_1,\delta_2)$}  && \multirow{2}{*}{Observable} & \multicolumn{3}{c}{Current ($\Lambda$ in TeV)} & \multicolumn{3}{c}{Future ($\Lambda$ in TeV)}  \\
      &&         & Constraint & $\Lambda_{\text{T}}\,(\Lambda_{\text{T}\chi})$ & $\Lambda_{\text{L}}\,(\Lambda_{\text{L}\chi})$ & Constraint & $\Lambda_{\text{T}}$ & $\Lambda_{\text{L}}$ \\\midrule
      \multirow{1}{*}{$\mathbf{2(0,1)}$} & $e\mu^{\dag}$ & $\text{BR}(\mu\to e\gamma)$ & $1.5\times10^{-13}$\cite{MEGII:2025gzr} & $3000$\,($73$) & $3000$\,($73$) & $6\times10^{-14}$\cite{MEGII:2021fah} & $3900$ & $3900$\\
                               &               & $\text{BR}(\mu\to ee\bar{e})$ & $1.0 \times 10^{-12}$\cite{SINDRUM:1987nra} & $500$\,($290$) & $520$\,($23$) & $10^{-16}$\cite{Hesketh:2022wgw} & $5000$ & $5200$\\
                               &               & $\text{CR}(\mu\,{\rm Au} \to e\,{\rm Au})$ & $7\times10^{-13}$\cite{SINDRUMII:2006dvw} & $1800$\,($1800$) & $430$\,($140$) & -- & -- & -- \\            
                               &               & $\text{CR}(\mu\,{\rm Al} \to e\,{\rm Al})$ & -- & -- & -- & $6\times10^{-17}$\cite{Mu2e:2014fns,COMET:2018auw} & $15000$ & $4000$ \\                               
                                    & $ee\mu^{\dag}\mu^{\dag}$ & $\text{P}(\rm M\rightarrow \rm \overline{M})$ & $8.2\times10^{-11}$\cite{Willmann:1998gd} & $9.5$ & $0.76$ & $10^{-13}$\cite{Bai:2024skk} & $51$ & $4.1$ \\
\midrule
      \multirow{1}{*}{$\mathbf{2(1,0)}$} & $\mu\tau^{\dag}$ & $\text{BR}(\tau\to \mu\gamma)$ & $4.2\times10^{-8}$\cite{Belle:2021ysv} & $20$\,($2.1$) & $20$\,($2.1$) & $6.9\times10^{-9}$\cite{Belle-II:2022cgf} & $32$ & $32$\\
                               &                  & $\text{BR}(\tau\to \mu\rho)$ & $1.7\times10^{-8}$\cite{Belle:2023ziz} & $21$\,($20$) & $5.9$\,($1.6$) & $5.5\times10^{-10}$\cite{Belle-II:2022cgf} & $49$ & $14$\\
                               &                  & $\text{BR}(\tau\to \mu\phi)$ & $2.3\times10^{-8}$\cite{Belle:2023ziz} & $14$\,($14$) & $3.2$\,($1.1$) & $8.4\times10^{-10}$\cite{Belle-II:2022cgf} & $33$ & $7.3$\\
                               &                  & $\text{BR}(\tau\to \mu\pi)$ & $1.1\times10^{-7}$\cite{BaBar:2006jhm} & $7.8$ & $0.62$ & $7.1\times10^{-10}$\cite{Belle-II:2022cgf} & $28$ & $2.2$\\
                               &                  & $\text{BR}(\tau\to \mu K)$ & $2.3\times10^{-8}$\cite{Belle:2010rxj} & $14$ & $1.1$ & $4.0\times10^{-10}$\cite{Belle-II:2022cgf} & $39$ & $3.1$ \\
                               &                  & $\text{BR}(\tau\to \mu\mu\bar{\mu})$ & $1.9\times10^{-8}$\cite{Belle-II:2024sce} & $16$\,($16$) & $5.3$\,($1.3$) & $3.6\times10^{-10}$\cite{Belle-II:2022cgf} & $42$ & $14$ \\
                               &                  & $\text{BR}(\tau\to \mu e\bar{e})$ & $1.8\times10^{-8}$\cite{Hayasaka:2010np} & $15$\,($15$) & $5.3$\,($1.2$) & $2.9\times10^{-10}$\cite{Belle-II:2022cgf} & $42$ & $15$ \\
                                    & $ee\mu^{\dag}\mu^{\dag}$ & $\text{P}({\rm M}\rightarrow \overline{\rm M})$ & $8.2\times10^{-11}$\cite{Willmann:1998gd} & $9.5$ & $0.76$ & $10^{-13}$\cite{Bai:2024skk} & $51$ & $4.1$ \\
                                    & $ee\mu^{\dag}\tau^\dag$ & $\text{BR}(\tau\to ee\bar{\mu})$ & $1.5\times10^{-8}$\cite{Hayasaka:2010np} & $16$ & $1.3$ & $2.3\times10^{-10}$\cite{Belle-II:2022cgf} & $45$ & $3.6$ \\
\midrule
      \multirow{1}{*}{$\mathbf{2(1,1)}$} & $e\tau^{\dag}$ & $\text{BR}(\tau\to e\gamma)$ & $3.3\times10^{-8}$\cite{BaBar:2009hkt} & $22$\,($2.2$) & $22$\,($2.2$) & $9.0\times10^{-9}$\cite{Belle-II:2022cgf} & $30$ & $30$\\
                               &                  & $\text{BR}(\tau\to e\rho)$ & $2.2\times10^{-8}$\cite{Belle:2023ziz} & $20$\,($19$) & $5.6$\,($1.5$) & $3.8\times10^{-10}$\cite{Belle-II:2022cgf} & $54$ & $15$\\
                               &                  & $\text{BR}(\tau\to e\phi)$ & $2.0\times10^{-8}$\cite{Belle:2023ziz} & $15$\,($15$) & $3.4$\,($1.2$) & $7.4\times10^{-10}$\cite{Belle-II:2022cgf} & $35$ & $7.7$\\
                               &                  & $\text{BR}(\tau\to e\pi)$ & $8.0\times10^{-8}$\cite{Belle:2007cio} & $8.5$ & $0.67$ & $7.3\times10^{-10}$\cite{Belle-II:2022cgf} & $27$ & $2.2$\\
                               &                  & $\text{BR}(\tau\to e K)$ & $2.6\times10^{-8}$\cite{Belle:2010rxj} & $14$ & $1.1$ & $4.0\times10^{-10}$\cite{Belle-II:2022cgf} & $40$ & $3.2$\\
                               &                  & $\text{BR}(\tau\to ee\bar{e})$ & $2.7\times10^{-8}$\cite{Hayasaka:2010np} & $15$\,($15$) & $7.3$\,($1.2$) & $4.7\times10^{-10}$\cite{Belle-II:2022cgf} & $40$ & $20$ \\
                               &                  & $\text{BR}(\tau\to e \mu\bar{\mu})$ & $2.7\times10^{-8}$\cite{Hayasaka:2010np} & $14$\,($14$) & $7.3$\,($1.1$) & $4.5\times10^{-10}$\cite{Belle-II:2022cgf} & $38$ & $20$ \\
                                    & $ee\mu^{\dag}\mu^{\dag}$ & $\text{P}({\rm M}\rightarrow \overline{\rm M})$ & $8.2\times10^{-11}$\cite{Willmann:1998gd} & $9.5$ & $0.76$ & $10^{-13}$\cite{Bai:2024skk} & $51$ & $4.1$\\
                                    & $e\mu^\dag\mu^{\dag}\tau$ & $\text{BR}(\tau\to \mu\mu\bar{e})$ & $1.7\times10^{-8}$\cite{Hayasaka:2010np} & $16$ & $1.2$ & $2.6\times10^{-10}$\cite{Belle-II:2022cgf} & $44$ & $3.5$ \\
\midrule
      \multirow{1}{*}{$\mathbf{3(1,1)}$} & $e\mu^{\dag}\mu^{\dag}\tau$ & $\text{BR}(\tau\to \mu\mu\bar{e})$ & $1.7\times10^{-8}$\cite{Hayasaka:2010np} & $16$ & $1.2$ & $2.6\times10^{-10}$\cite{Belle-II:2022cgf} & $44$ & $3.5$\\
                            & $ee\mu^{\dag}\tau^{\dag}$ & $\text{BR}(\tau\to ee\bar{\mu})$ & $1.5\times10^{-8}$\cite{Hayasaka:2010np} & $16$ & $1.3$ & $2.3\times10^{-10}$\cite{Belle-II:2022cgf} & $45$ & $3.6$\\
\midrule
      \multirow{1}{*}{$\mathbf{4(1,1)}$} & $e\mu^{\dag}\mu^{\dag}\tau$ & $\text{BR}(\tau\to \mu\mu\bar{e})$ & $1.7\times10^{-8}$\cite{Hayasaka:2010np} & $16$ & $1.2$ & $2.6\times10^{-10}$\cite{Belle-II:2022cgf} & $44$ & $3.5$ \\
\midrule
   \multirow{1}{*}{$\mathbf{4(2,3)}$} & $ee\mu^{\dag}\mu^{\dag}$ & $\text{P}({\rm M}\rightarrow \overline{\rm M})$ & $8.2\times10^{-11}$\cite{Willmann:1998gd} & $9.5$ & $0.76$ & $10^{-13}$\cite{Bai:2024skk} & $51$ & $4.1$\\
\midrule
   \multirow{1}{*}{$\mathbf{4(3,2)}$} & $ee\mu^{\dag}\tau^{\dag}$ & $\text{BR}(\tau\to ee\bar{\mu})$ & $1.5\times10^{-8}$\cite{Hayasaka:2010np} & $16$ & $1.3$ & $2.3\times10^{-10}$\cite{Belle-II:2022cgf} & $45$ & $3.6$\\
\bottomrule
   \end{tabular}}
   \caption{
   Numerical values of the current lower bounds on and future sensitivities to the new physics scale $\Lambda$. All experimental limits, current and expected, are given at 
$90\%$ C.L.. Three different scenarios are considered: one with tree-level contributions, $C_x=1$, except the WCs of the dipole operators, $C_{d}=e/(16\pi^2)$, leading to the limits denoted as $\Lambda_{\rm T}$; one with loop-level contributions, $C_x=1/(16\pi^2)$, resulting in $\Lambda_{\rm L}$; a third one with chirally suppressed WCs of the dipole operators, $C_{d} = \sqrt{2} \, m_\ell \, e/(16\pi^2 v)$, 
   in each of the former two scenarios, corresponding to $\Lambda_{{\rm T}\chi}$ and $\Lambda_{{\rm L}\chi}$, respectively. 
}   
   \label{tbl:scaleData}   
\end{table}

\clearpage

\begin{table}[ht!]
   \centering
   \renewcommand\arraystretch{1.1}
        \setlength{\tabcolsep}{5pt}
   \begin{tabular}{lrlcccc}
   \toprule
\multirow{2}{*}{$N(\delta_1,\delta_2)$}&& \multirow{2}{*}{Observable} & \multicolumn{2}{c}{Current ($\Lambda$ in TeV)} & \multicolumn{2}{c}{Future ($\Lambda$ in TeV)}  \\
&&         & Constraint & $\Lambda_{\text{T}}$ & Constraint & $\Lambda_{\text{T}}$ \\\midrule
    $\mathbf{2(a,b)}^\ddag$, $\mathbf{4(3,2)}$&$\mu\mu\tau^{\dag}\tau^{\dag}$ & $\sigma(\mu^+\mu^+\to\tau^+\tau^+)$ & -- & -- & $0.3$~fb~\cite{Hamada:2022mua} & 30 \\
    && ${\rm BR}(Z\to \tau\tau\bar{\mu}\bar{\mu})$ & $2\times 10^{-3}$~\cite{Heeck:2024uiz} & 0.001 &  $10^{-12}$~\cite{Heeck:2024uiz} &  0.25 \\
    $\mathbf{2(a,b)}^\ddag$, $\mathbf{4(1,1)}$&$ee\tau^{\dag}\tau^{\dag}$ 
    & ${\rm BR}(Z\to \tau\tau\bar{e}\bar{e})$ & $2\times 10^{-3}$~\cite{Heeck:2024uiz} &  0.001 & $10^{-12}$~\cite{Heeck:2024uiz} & 0.25 \\
    $\mathbf{2(0,1)}$, $\mathbf{3(1,1)}$, $\mathbf{4(2,3)}$&$e\mu\tau^{\dag}\tau^{\dag}$ & ${\rm BR}(Z\to \tau\tau\bar{e}\bar{\mu})$ & $2\times 10^{-3}$~\cite{Heeck:2024uiz} & 0.001 & $10^{-12}$~\cite{Heeck:2024uiz} & 0.21 \\
\bottomrule
   \end{tabular}
   \flushleft
   $^\ddag$\,$\mathbf{2(a,b)}$ stands for $\mathbf{2(0,1)}$, $\mathbf{2(1,0)}$ and $\mathbf{2(1,1)}$.
   \caption{Compilation of current limits on and future sensitivities to the new physics scale $\Lambda$ arising from high-energy probes of cLFV physics. All calculations assume the scenario with tree-level contributions, see~\cref{sec:collider} for details.
   The notation follows that of~\Cref{tbl:scaleData}.
}
   \label{tbl:scaleDataCollider}   
\end{table}

\newpage

\section{Constraints on \mathversion{bold}\texorpdfstring{$\Lambda$}{Lambda}\mathversion{normal}
for remaining flavour charge assignments}
\label{app:scale2}
\begin{figure}[t!]
   \centering
   \includegraphics[width=\textwidth]{"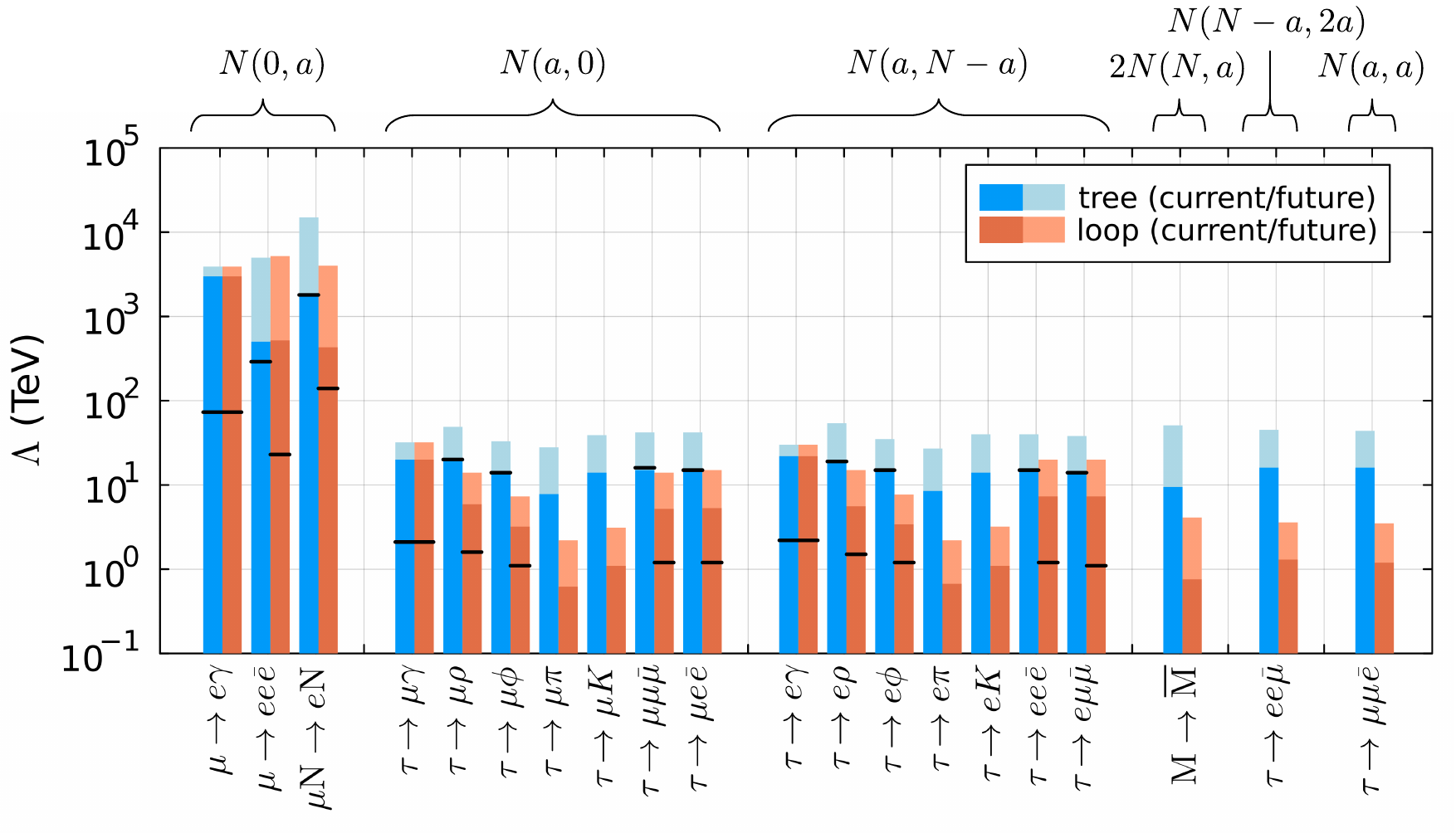"}
   \caption{Current 
    lower bounds on and future sensitivities to the new physics scale $\Lambda$ (in~$\rm TeV$), based on experimental limits for various processes at $90\%\,\rm C.L.$, for flavour charge assignments listed in the table at the bottom of Figure~\ref{fig:rotary}.
  Conventions are the same as in Figure~\ref{fig:scalebar}.
  }
   \label{fig:scalebar2}
\end{figure}

We present for the flavour charge assignments not appearing in Figure~\ref{fig:rotary} the tables analogous to~\cref{tbl:scaleData,tbl:scaleDataCollider} (see~\cref{tbl:scaleData2,tbl:scaleDataCollider2}) as well as the figure corresponding to Figure~\ref{fig:scalebar} (see Figure~\ref{fig:scalebar2}).

\begin{table}[ht!]
   \centering
   \renewcommand\arraystretch{1.}
        \setlength{\tabcolsep}{2pt}
   \resizebox{\linewidth}{!}{\begin{tabular}{crlcccccc}
   \toprule
     \multirow{2}{*}{$N(\delta_1,\delta_2)$}  && \multirow{2}{*}{Observable} & \multicolumn{3}{c}{Current ($\Lambda$ in TeV)} & \multicolumn{3}{c}{Future ($\Lambda$ in TeV)}  \\
      &&         & Constraint & $\Lambda_{\text{T}}\,(\Lambda_{\text{T}\chi})$ & $\Lambda_{\text{L}}\,(\Lambda_{\text{L}\chi})$ & Constraint & $\Lambda_{\text{T}}$ & $\Lambda_{\text{L}}$ \\\midrule
      \multirow{1}{*}{$N(0,a)$} & $e\mu^{\dag}$ & $\text{BR}(\mu\to e\gamma)$ & $1.5\times10^{-13}$\cite{MEGII:2025gzr} & $3000$\,($73$) & $3000$\,($73$) & $6\times10^{-14}$\cite{MEGII:2021fah} & $3900$ & $3900$\\
                               &               & $\text{BR}(\mu\to ee\bar{e})$ & $1.0 \times 10^{-12}$\cite{SINDRUM:1987nra} & $500$\,($290$) & $520$\,($23$) & $10^{-16}$\cite{Hesketh:2022wgw} & $5000$ & $5200$\\
                               &               & $\text{CR}(\mu\,{\rm Au} \to e\,{\rm Au})$ & $7\times10^{-13}$\cite{SINDRUMII:2006dvw} & $1800$\,($1800$) & $430$\,($140$) & -- & -- & -- \\            
                               &               & $\text{CR}(\mu\,{\rm Al} \to e\,{\rm Al})$ & -- & -- & -- & $6\times10^{-17}$\cite{Mu2e:2014fns,COMET:2018auw} & $15000$ & $4000$ \\                               
\midrule
      \multirow{1}{*}{$N(a,0)$} & $\mu\tau^{\dag}$ & $\text{BR}(\tau\to \mu\gamma)$ & $4.2\times10^{-8}$\cite{Belle:2021ysv} & $20$\,($2.1$) & $20$\,($2.1$) & $6.9\times10^{-9}$\cite{Belle-II:2022cgf} & $32$ & $32$\\
                               &                  & $\text{BR}(\tau\to \mu\rho)$ & $1.7\times10^{-8}$\cite{Belle:2023ziz} & $21$\,($20$) & $5.9$\,($1.6$) & $5.5\times10^{-10}$\cite{Belle-II:2022cgf} & $49$ & $14$\\
                               &                  & $\text{BR}(\tau\to \mu\phi)$ & $2.3\times10^{-8}$\cite{Belle:2023ziz} & $14$\,($14$) & $3.2$\,($1.1$) & $8.4\times10^{-10}$\cite{Belle-II:2022cgf} & $33$ & $7.3$\\
                               &                  & $\text{BR}(\tau\to \mu\pi)$ & $1.1\times10^{-7}$\cite{BaBar:2006jhm} & $7.8$ & $0.62$ & $7.1\times10^{-10}$\cite{Belle-II:2022cgf} & $28$ & $2.2$\\
                               &                  & $\text{BR}(\tau\to \mu K)$ & $2.3\times10^{-8}$\cite{Belle:2010rxj} & $14$ & $1.1$ & $4.0\times10^{-10}$\cite{Belle-II:2022cgf} & $39$ & $3.1$ \\
                               &                  & $\text{BR}(\tau\to \mu\mu\bar{\mu})$ & $1.9\times10^{-8}$\cite{Belle-II:2024sce} & $16$\,($16$) & $5.3$\,($1.3$) & $3.6\times10^{-10}$\cite{Belle-II:2022cgf} & $42$ & $14$ \\
                               &                  & $\text{BR}(\tau\to \mu e\bar{e})$ & $1.8\times10^{-8}$\cite{Hayasaka:2010np} & $15$\,($15$) & $5.3$\,($1.2$) & $2.9\times10^{-10}$\cite{Belle-II:2022cgf} & $42$ & $15$ \\
\midrule
      \multirow{1}{*}{$N(a,N-a)$} & $e\tau^{\dag}$ & $\text{BR}(\tau\to e\gamma)$ & $3.3\times10^{-8}$\cite{BaBar:2009hkt} & $22$\,($2.2$) & $22$\,($2.2$) & $9.0\times10^{-9}$\cite{Belle-II:2022cgf} & $30$ & $30$\\
                               &                  & $\text{BR}(\tau\to e\rho)$ & $2.2\times10^{-8}$\cite{Belle:2023ziz} & $20$\,($19$) & $5.6$\,($1.5$) & $3.8\times10^{-10}$\cite{Belle-II:2022cgf} & $54$ & $15$\\
                               &                  & $\text{BR}(\tau\to e\phi)$ & $2.0\times10^{-8}$\cite{Belle:2023ziz} & $15$\,($15$) & $3.4$\,($1.2$) & $7.4\times10^{-10}$\cite{Belle-II:2022cgf} & $35$ & $7.7$\\
                               &                  & $\text{BR}(\tau\to e\pi)$ & $8.0\times10^{-8}$\cite{Belle:2007cio} & $8.5$ & $0.67$ & $7.3\times10^{-10}$\cite{Belle-II:2022cgf} & $27$ & $2.2$\\
                               &                  & $\text{BR}(\tau\to e K)$ & $2.6\times10^{-8}$\cite{Belle:2010rxj} & $14$ & $1.1$ & $4.0\times10^{-10}$\cite{Belle-II:2022cgf} & $40$ & $3.2$\\
                               &                  & $\text{BR}(\tau\to ee\bar{e})$ & $2.7\times10^{-8}$\cite{Hayasaka:2010np} & $15$\,($15$) & $7.3$\,($1.2$) & $4.7\times10^{-10}$\cite{Belle-II:2022cgf} & $40$ & $20$ \\
                               &                  & $\text{BR}(\tau\to e \mu\bar{\mu})$ & $2.7\times10^{-8}$\cite{Hayasaka:2010np} & $14$\,($14$) & $7.3$\,($1.1$) & $4.5\times10^{-10}$\cite{Belle-II:2022cgf} & $38$ & $20$ \\
\midrule
   \multirow{1}{*}{$2N(N,a)$} & $ee\mu^{\dag}\mu^{\dag}$ & $\text{P}({\rm M}\rightarrow \overline{\rm M})$ & $8.2\times10^{-11}$\cite{Willmann:1998gd} & $9.5$ & $0.76$ & $10^{-13}$\cite{Bai:2024skk} & $51$ & $4.1$\\
\midrule
   \multirow{1}{*}{$N(N-a,2a)$} & $ee\mu^{\dag}\tau^{\dag}$ & $\text{BR}(\tau\to ee\bar{\mu})$ & $1.5\times10^{-8}$\cite{Hayasaka:2010np} & $16$ & $1.3$ & $2.3\times10^{-10}$\cite{Belle-II:2022cgf} & $45$ & $3.6$\\
\midrule
      \multirow{1}{*}{$N(a,a)$} & $e\mu^{\dag}\mu^{\dag}\tau$ & $\text{BR}(\tau\to \mu\mu\bar{e})$ & $1.7\times10^{-8}$\cite{Hayasaka:2010np} & $16$ & $1.2$ & $2.6\times10^{-10}$\cite{Belle-II:2022cgf} & $44$ & $3.5$ \\
\bottomrule
   \end{tabular}}
   \caption{
   Numerical values of the current lower bounds on and future sensitivities to the new physics scale $\Lambda$ for flavour charge assignments listed in the table at the bottom of Figure~\ref{fig:rotary}. Conventions are the same as in Table~\ref{tbl:scaleData}.}   
   \label{tbl:scaleData2}   
\end{table}

\begin{table}[ht!]
   \centering
   \renewcommand\arraystretch{1.1}
        \setlength{\tabcolsep}{5pt}
   \begin{tabular}{lrlcccc}
   \toprule
\multirow{2}{*}{$N(\delta_1,\delta_2)$}&& \multirow{2}{*}{Observable} & \multicolumn{2}{c}{Current ($\Lambda$ in TeV)} & \multicolumn{2}{c}{Future ($\Lambda$ in TeV)}  \\
&&         & Constraint & $\Lambda_{\text{T}}$ & Constraint & $\Lambda_{\text{T}}$ \\\midrule
    $2N(a,N)$
    &$\mu\mu\tau^{\dag}\tau^{\dag}$ & $\sigma(\mu^+\mu^+\to\tau^+\tau^+)$ & -- & -- & $0.3$~fb~\cite{Hamada:2022mua} & 30 \\
    && ${\rm BR}(Z\to \tau\tau\bar{\mu}\bar{\mu})$ & $2\times 10^{-3}$~\cite{Heeck:2024uiz} & 0.001 &  $10^{-12}$~\cite{Heeck:2024uiz} &  0.25 \\
    $2N(a,N-a)$&$ee\tau^{\dag}\tau^{\dag}$ 
    & ${\rm BR}(Z\to \tau\tau\bar{e}\bar{e})$ & $2\times 10^{-3}$~\cite{Heeck:2024uiz} &  0.001 & $10^{-12}$~\cite{Heeck:2024uiz} & 0.25 \\
    $N(2a,N-a)$&$e\mu\tau^{\dag}\tau^{\dag}$ & ${\rm BR}(Z\to \tau\tau\bar{e}\bar{\mu})$ & $2\times 10^{-3}$~\cite{Heeck:2024uiz} & 0.001 & $10^{-12}$~\cite{Heeck:2024uiz} & 0.21 \\
\bottomrule
   \end{tabular}
   \caption{Compilation of current limits on and future sensitivities to the new physics scale $\Lambda$ arising from high-energy probes of cLFV physics for flavour charge assignments listed in the table at the bottom of Figure~\ref{fig:rotary}. Conventions are the same as in Table~\ref{tbl:scaleDataCollider}.}
   \label{tbl:scaleDataCollider2}   
\end{table}

\bibliographystyle{utphys}
\bibliography{refsv2.bib}

\end{document}